\documentclass[aps,superscriptaddress,prd,
onecolumn,
floatfix, 
nofootinbib,
amsmath,amssymb,amsfonts,longbibliography]{revtex4-2}

\usepackage{mathrsfs,amsmath}
\usepackage[pdftex]{graphicx}
\usepackage[dvipsnames]{xcolor}
\usepackage{float}
\usepackage{physics}
\usepackage{comment}
\usepackage[normalem]{ulem}
\usepackage{caption}
\usepackage[colorlinks,linkcolor=blue,anchorcolor=violet,citecolor=red]{hyperref}
\newcommand{\souteq}[1]{\hbox{}}
\newcommand{\red}[1]{\textcolor{red}{#1}}

\newcommand{\green}[1]{\textcolor{Green}{#1}}

\begin{document}

\title{
Enhancement of quantum gravity signal in an optomechanical experiment
}

\author{Youka Kaku}
\email{kaku.yuka.g4@s.mail.nagoya-u.ac.jp}
\affiliation{Department of Physics, Graduate School of Science, Nagoya University, Chikusa, Nagoya 464-8602, Japan}
\author{Tomohiro Fujita}
\email{tomofuji@aoni.waseda.jp}
\affiliation{Waseda Institute for Advanced Study, Waseda University,
1-6-1 Nishi-Waseda, Shinjuku, Tokyo 169-8050, Japan}
\affiliation{Research Center for the Early Universe, The University of Tokyo, Bunkyo, Tokyo 113-0033, Japan}
\author{Akira Matsumura}
\email{matsumura.akira@phys.kyushu-u.ac.jp}
\affiliation{Department of Physics, Kyushu University, Fukuoka, 819-0395, Japan}

\date{\today}

\begin{abstract}

No experimental evidence of the quantum nature of gravity has been observed yet and a realistic setup with improved sensitivity is eagerly awaited. 
We find two effects, which can substantially enhance the signal of gravity-induced quantum entanglement, by examining an optomechanical system in which two oscillators gravitationally couple and one
composes an optical cavity.
The first effect comes from a higher-order term of the optomechanical interaction and generates the signal at the first order of the gravitational coupling in contrast to the second order results in previous works.
The second effect is the resonance between the two oscillators. If their frequencies are close enough, the weak gravitational coupling effectively strengthens.
Combining these two effects, the signal in the interference visibility could be amplified by a factor of $10^{24}$ for our optimistic parameters. 
The two effects would be useful in seeking feasible experimental setups to probe quantum gravity signals.
\end{abstract}

\maketitle

\section{Introduction}

The construction of a quantum gravity theory poses a fundamental challenge in theoretical physics \cite{Kiefer2006,Woodard2009}. 
One of the main difficulties stems from the lack of sufficient experimental evidence to investigate quantum gravity.
As a first step addressing this issue, Feynman proposed a thought experiment to observe a probe system evolving under a quantum superposition of gravitational fields \cite{Rickles2011}. 
This idea inspired the investigations of quantum coherent phenomena on a low-energy tabletop experiment.
A novel proposal is often commonly referred to as the Bose et al.-Matletto-Vedral (BMV) proposal 
\cite{Bose2017,Marletto2017}.
In Ref.\cite{Bose2017,Marletto2017}, the authors considered a scenario where two quantum masses, initially in a non-entangled state and each in a spatial superposition, interact only through Newtonian gravity.
They concluded that the entanglement between the masses is generated by the gravitational interaction and that such a phenomenon indicates the quantum coherent behavior of gravity.  
Stimulated by this statement, there are many experimental proposals based on matter-wave interferometers \cite{Nguyen2020, Chevalier2020, vandeKamp2020, Toros2021, Miki2021a,Tilly2021}, mechanical oscillator model \cite{Qvafort2020, Krisnanda2020}, optomechanical systems \cite{Balushi2018, Miao2020, Wan2017, Matsumura2020, Miki2021b} and their hybrid model \cite{Carney2021a, Carney2022, Streltsov2022, Pedernales2021, Matsumura2021b}.
Also, the theoretical aspects of gravity-induced entanglement have been studied. 
In Refs.\cite{Belenchia2018, Danielson2022, Marshman2020, Hidaka2022, Sugiyama2022, Christodoulou2023}, the entanglement due to Newtonian gravity was shown to be consistent with quantum field theoretical description.
On the other hand, it was discussed that such a Newtonian entanglement does not directly lead to the quantization of the gravitational field \cite{Hall2018, Anastopoulos2022}. 
In this context, it may also be interesting to verify the entanglement due to gravity in a relativistic regime \cite{Christodoulou2019, Bose2022, Biswas2022, Kaku2022}.

The above major trends pave the way to uncovering the quantum aspects of gravity. 
Recent advancements in optomechanics \cite{Schmole2016, Matsumoto2019, Matsumoto2020, Michimura2020, Miki2023, Sugiyama2023, Shichijo2023} further encourage us to investigate the quantum signal of gravity in an optomechanical setup. 
In this direction, Balushi et al. proposed such a setup involving two mechanical oscillators interacting through Newtonian gravity, each of which is coupled to an optomechanical interferometer \cite{Balushi2018}.
The authors demonstrated that the gravitational interaction between the quantum oscillators induces an effective frequency shift of photons within the interferometer, and this results in a dephasing of photon interference visibility. 
In Ref.\cite{Matsumura2020}, the entanglement generation due to gravity in the setup was analyzed in an exact non-perturbative manner. 

Despite various efforts to realize quantum gravity experiments, no experimental evidence of quantum gravity has been observed to date. 
In this paper, we present an amplification of the quantum signal of gravity in an optomechanical setup.
Inspired by the work of Balushi et al. \cite{Balushi2018}, we consider a hybrid system consisting of two oscillators and one optomechanical interferometer. 
In this setup, the two oscillators interact with each other by gravity, and one of the oscillators is coupled to a single photon in the interferometer via an optomechanical interaction.
In Ref.\cite{Balushi2018}, they considered the leading order of an optomechanical interaction and demonstrated that the modification of the photon interference visibility is of the second order of gravitational coupling.
In comparison, we treat up to the sub-leading order of the optomechanical interaction. 
As a result, we find that the visibility deviates by the first order of gravitational coupling.
In other words, the large signal of gravity can be observed in the experiment accessible to the higher-order optomechanical coupling.
We further investigate how the resonance of the two oscillators affects the gravitational deviation of the visibility. 
It is then demonstrated that such a deviation
is amplified with the inverse of a frequency difference between two oscillators, provided that quantum coherence can be maintained in the system for a sufficiently long time.
We finally evaluate the entanglement due to gravity in this system. 
Focusing on the resonance effect, we discuss the relationship between gravity-induced entanglement and the gravitational deviation of visibility. 

This paper is organized as follows: The setup and the Hamiltonian are introduced in section \ref{sec:setup}. Then, we investigate the time evolution of the system in section \ref{sec:evolution}. In section \ref{sec:visibility}, we show the visibility of single-photon interference in the optomechanical setup and discuss that the gravitational effect appears as a lower order of the gravitational coupling constant compared to Ref.\cite{Balushi2018}. We present numerical results of the visibility in section \ref{sec:numerical results}. In Section \ref{sec:resonance}, we examine the resonance effect on the visibility.
We also estimate quantum entanglement generated by gravity in section \ref{sec:entanglement}, and clarify the relationship between the entanglement generation and the gravitational deviation in the visibility. Finally, we summarize the paper in section \ref{sec:conclusion}.

\section{The setup and Hamiltonian
}\label{sec:setup}

\begin{figure}[htbp]
    \centering
    \includegraphics[width=0.6\linewidth]{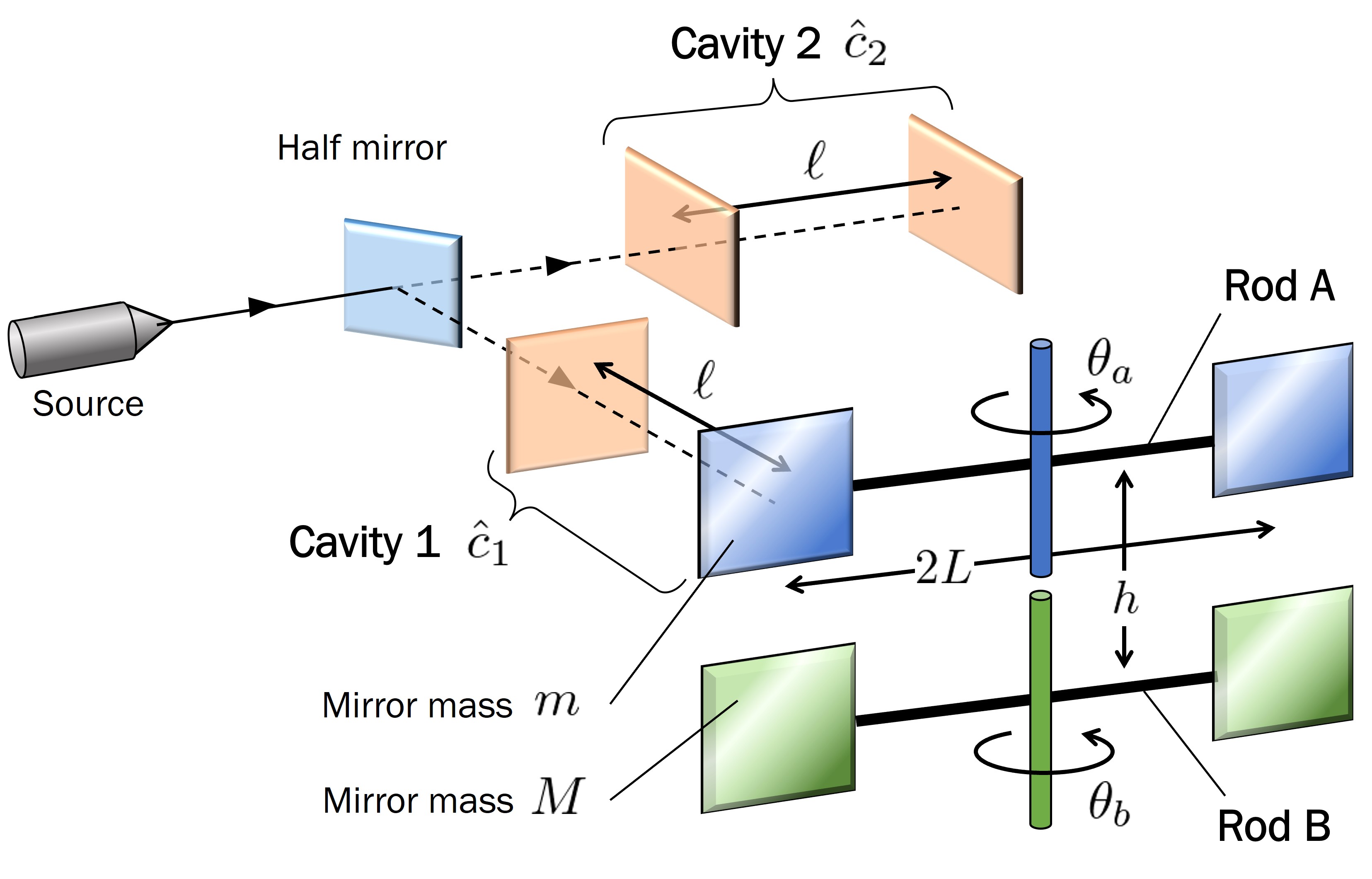}
    \caption{Our setup with two optical cavities and two micro mechanical rods. A single photon emitted by the source is prepared in a quantum superposition state in cavity 1 and 2 by a half mirror. Rod A 
    and cavity 1 form an optomechanical system. 
    The photon in cavity 1 and the mirror of mass $m$ attached to rod A interact with each other. The mirrors of rod B are coupled to the mirrors of rod A only through gravity. Quantum entanglement between rod B and the other system of the setup (i.e. rod A and the photon in the cavities) mediated by the gravitational coupling could be measured by the change of the interference visibility of the photons.}
    \label{fig:setup}
\end{figure}

Let us consider a cavity optomechanical system to detect the quantum feature of gravity.
Fig.~\ref{fig:setup} illustrates an experimental setup with a pair of cavities and two micro mechanical rods of length $2L$.
The two rods are suspended by independent center bars with a vertical separation $h$ and can oscillate in a horizontal plane.
A single photon emitted by the source passes through the half mirror and then is in a superposition of the state being in cavity 1 and in cavity 2.
Here, the annihilation and creation operators of the photon in cavities 1 and 2 are represented as $\{\hat c_1,~\hat c_1^\dagger\}$ and $\{\hat c_2,~\hat c_2^\dagger\}$, respectively. 
The photon in the cavity 1 pushes the mirror of mass $m$ at the left end of rod A and interacts with the mechanical mode of oscillating rod A. 
The oscillation of rod A is characterized by its angular position and momentum operator $\hat \theta_a, \hat p_a$. 
Its moment of inertia and angular frequency are defined as $I_a=2mL^2$ and $\Omega_a$, respectively.
Rod B with the mirror mass $M$ interacts with Rod A only through gravity. 
Similarly, the oscillation of rod B is characterized by $\hat \theta_b,~\hat p_b$, and its moment of inertia and frequency are given by $I_b=2ML^2,~\Omega_b$.
This setup is based on the system proposed in Ref.\cite{Balushi2018}.
They considered another set of cavities interacting with rod B, which is removed in our setup for simplicity. 
To analytically solve the dynamics of this system, we assume that the vertical separation of the rods is much smaller than their length $2L\gg h$, and the oscillations of rods A and B are small $\theta_a,~\theta_b\ll1$. 

Let us consider the optomechanical coupling between the photon in cavity 1 and rod A by taking a higher-order correction into account.
When the mirror $m$ is in the original position $\theta_a=0$, the photon frequency of the cavity mode would be
\begin{align}
    \omega_c=\frac{\pi c\, \mathfrak{n}}{\ell}.
\end{align}
where $\ell$ is the original cavity length, $c$ is the speed of light and $\mathfrak{n}$ is an integer. 
When a photon enters cavity 1 and pushes the mirror $m$, 
the frequency of the cavity mode is modified as 
\begin{align}
    \omega_c'=\frac{\pi c\, \mathfrak{n}}{\ell+L\sin\theta_a}
    \approx \omega_c\left(1-\frac{L}{\ell}\theta_a+\frac{L^2}{\ell^2}\theta_a^2\right).
    \label{eq:thetam_expansion}
\end{align}
Here we include the second order of 
$\theta_a$, which was neglected in the previous works \cite{Balushi2018,Matsumura2020}. 
This second-order correction might appear to be a sub-leading effect of the optomechanical coupling between the photon and rod A, which slightly distorts the harmonic oscillator potential of rod A.
However, we will show that this contribution 
has a significant impact on the signal of the quantum nature of gravity.

We organize the total Hamiltonian up to the second order of $\theta_a$ as
\begin{align}
    \hat H
    &=\hbar\omega_c'\hat c_1^\dagger\hat c_1
    +\hbar\omega_c\hat c_2^\dagger\hat c_2
    +\frac{1}{2 I_a}\hat p_a^2+\frac{1}{2}I_a \Omega_a^2\hat \theta_a^2
    +\frac{1}{2 I_b}\hat p_b^2+\frac{1}{2}I_b \Omega_b^2\hat \theta_b^2
    +\frac{GmML^2}{h^3}\left(\hat\theta_a^2+\hat\theta_b^2-2\hat\theta_a\hat\theta_b\right)
    \notag\\
    &\approx \hbar\omega_c\hat c_1^\dagger\hat c_1
    +\hbar\omega_c\hat c_2^\dagger\hat c_2
    +\frac{1}{2 I_a}\hat p_a^2
    +\frac{I_a}{2}\left(\Omega_a^2+\frac{GM}{h^3}+\frac{\hbar\omega_c}{m\ell^2}\hat c_1^\dagger\hat c_1\right)\hat \theta_a^2
    -\frac{\hbar\omega_c L}{\ell}\hat c_1^\dagger\hat c_1\hat\theta_a
    \notag\\
    &\hspace{150pt}+\frac{1}{2 I_b}\hat p_b^2
    +\frac{I_b}{2}\left(\Omega_b^2+\frac{Gm}{h^3}\right)\hat\theta_b^2
    +2\frac{GmML^2}{h^3}\hat\theta_a\hat\theta_b
    \notag\\
    &=\hbar\omega_c\hat c_1^\dagger\hat c_1+\hbar\omega_c\hat c_2^\dagger\hat c_2 +\sum_{n=0,1}\hat H_{a,n}|n\rangle_{c1\,c1}\langle n|+\hat H_b+\hat H_g,
    \label{eq:total_Hamiltonian}
\end{align}
where we plugged Eq.~\eqref{eq:thetam_expansion} in the second line. 
The last term in the first line denotes the gravitational interaction part, which is derived in Appendix.~\ref{apdx:Hamiltonian}. 
In the last line of Eq.\eqref{eq:total_Hamiltonian}, $\ket{n}_{c1}$ denotes an eigenstate of the photon number in cavity 1, $\hat c_1^\dagger\hat c_1$, and 
$\hat H_{a,n}$ is an effective Hamiltonian of rod A depending on the photon number $n$ inside cavity 1. 
This Hamiltonian containing the optomechanical coupling with the cavity photon is given as
\begin{align}
\label{eq:Hamiltonian a}
    &\hat H_{a,n}
    =\hbar\omega_{a,n}\left\{\hat a^\dagger_n \hat a_n-n\lambda_{n}(\hat a^\dagger_n+\hat a_n)+\frac{1}{2}\right\},\\
    &\hat a_n =\sqrt{\frac{I_a\omega_{a,n}}{2\hbar}}\hat \theta_a+\frac{i}{\sqrt{2I_a \omega_{a,n}\hbar}}\hat p_a,\quad
    \lambda_{n}=\left(\frac{\omega_{a,0}}{\omega_{a,n}}\right)^{3/2}\frac{\omega_c}{\omega_{a,0}}\sqrt{\frac{\hbar}{2I_a\omega_{a,0}}}\frac{L}{\ell},
\end{align}
where $\lambda_n$ denotes an optomechanical coupling constant, and the oscillation frequency of rod A depends on the photon number $n$ as
\begin{align}
\omega_{a,n}=\sqrt{\Omega_a^2+\frac{G M}{h^3}}\times \left\{
    \begin{array}{ll}
        \,1 & (n=0;\ \text{no photon in cavity 1}) \\
        \sqrt{1+\frac{2\hbar\omega_c}{I_a\omega_a^2}\frac{L^2}{\ell^2}}\quad & (n=1;\ \text{photon pressure distorts the potential})
    \end{array}
    \right.\,.
    \label{eq:omega01}
\end{align}
The effective Hamiltonian of rod B, $\hat{H}_b$, and the gravitational interaction term $\hat{H}_g$ in Eq.\eqref{eq:total_Hamiltonian} are defined as
\begin{align}
    \hat H_b=\hbar\omega_b \hat b^\dagger\hat b,\quad
    \hat b=\sqrt{\frac{I_b\omega_{b}}{2\hbar}}\hat \theta_b+\frac{i}{\sqrt{2I_b \omega_{b}\hbar}}\hat p_b,\quad
    \omega_b=\sqrt{\Omega_b^2+\frac{G m}{h^3}},
\end{align}
and 
\begin{align}
    \hat H_g=-g \hbar\omega_{a,0} (\hat a^\dagger_0+\hat a_0)(\hat b^\dagger+\hat b),
    \quad
    g=\frac{G}{2h^3\omega_{a,0}}\sqrt{\frac{m M}{\omega_{a,0}\omega_b}},
\end{align}
respectively. 

As observed in Eq.\eqref{eq:omega01}, the oscillation frequency of rod A is shifted from $\Omega_a$ due to the gravitational interaction between the rods.
Moreover, the frequency also differs depending on the photon number $n$; If the photon hits the mirror of rod A ($n=1$), the mechanical potential of rod A is not only displaced by the $\mathcal{O}[\theta_a]$ term in Eq.~\eqref{eq:thetam_expansion}, but also distorted according to the $\mathcal{O}[\theta_a]^2$ term. The distortion of the potential is reinterpreted as the shift in the cavity mode frequency from $\omega_{a,0}$ to $\omega_{a,1}$. Remark that this frequency shift due to the optomechanical coupling was not considered in Ref.\cite{Balushi2018}, and this is the key novelty in our paper. 
In comparison to the previous works, the frequency ratio $\omega_{a,0}/\omega_{a,1}$ will be an important parameter as the new effect from the higher order contribution of $\theta_a$.

\section{The evolution of the system}\label{sec:evolution}

Now we are ready to solve the time evolution of our system. The initial state of the total system is prepared as a non-entangled state
\begin{align}
    |\psi(t=0)\rangle
    =\frac{1}{\sqrt{2}}\big(\ket{0}_{c1}\ket{1}_{c2}+\ket{1}_{c1}\ket{0}_{c2}\big)\otimes \ket{\alpha}_a\otimes \ket{\beta}_b.
    \label{eq:initial_state}
\end{align}
Here, $\ket{n}_{c1}\ket{n'}_{c2}$ denotes the photon state when $n$ number of photons enters cavity 1 and $n'$ number of photons enters cavity 2. 
$\ket{\alpha}_a$ is a coherent state of the mechanical mode of rod A for $\hat a_0$ (not for $\hat a_1$), and $\ket{\beta}_b$ is a coherent state of that of rod B.
Since the gravitational coupling is very small $g\ll 1$, 
we evaluate the evolved state by perturbation theory with respect to $g$.
Up to the first order of $g$,  we obtain
\begin{align}
    |\psi(t)\rangle&=e^{-i\hat H t/\hbar}|\psi(0)\rangle
    \notag\\
    &=\frac{e^{-i\omega_c t}}{\sqrt{2}}\sum_{n=0,1}
    \ket{n}_{c1}\ket{1-n}_{c2}
    e^{-i\left(\hat H_{a,n}+\hat H_b\right)t/\hbar}
    \left[1-\frac{i}{\hbar}\int^t_0 dt' \hat H^{I}_{g,n}(t')\right]
    \ket{\alpha}_a \ket{\beta}_b
    +\mathcal{O}(g^2),
    \notag\\
    &=\frac{e^{-i\omega_c t}}{\sqrt{2}}\sum_{n=0,1}
    \ket{n}_{c1}\ket{1-n}_{c2}
    \left[1+2ig\left(\hat{\mathcal{I}}_{n}(t)+n\hat{\mathcal{J}}(t)\right)\right]
    e^{-i\left(\hat H_{a,n}+\hat H_b\right)t/\hbar}\ket{\alpha}_a \ket{\beta}_b+\mathcal{O}(g^2),
    \label{eq:time_evolution}
\end{align}
where $\hat H^{I}_{g,n}(t)=e^{i(\hat H_{a,n}+\hat H_b)t/\hbar}\hat H_g e^{-i(\hat H_{a,n}+\hat H_b)t/\hbar}$ 
is the gravitational interaction in the interaction picture.
From the second line to the third line, we used 
$e^{-i\left(\hat H_{a,n}+\hat H_b\right)t/\hbar}\hat H^I_{g,n}(t')=\hat H^I_{g,n}(t'-t)e^{-i\left(\hat H_{a,n}+\hat H_b\right)t/\hbar}$ and performed the $t'$ integration, which yielded 
new Hermitian operators
\begin{align}
    &\hat{\mathcal{I}}_{n}(t)
    :=\sqrt{\frac{\omega_{a,0}^3}{\omega_{a,n}}}\left\{\frac{\sin[\omega_{n,+}t/2]}{\omega_{n,+}}\left(e^{-i\omega_{n,+}t/2}\hat a_{n}^\dagger\hat b^\dagger+e^{i\omega_{n,+}t/2}\hat a_{n}\hat b\right)
    +\frac{\sin[\omega_{n,-}t/2]}{\omega_{n,-}}\left(e^{-i\omega_{n,-}t/2}\hat a_{n}^\dagger\hat b+e^{i\omega_{n,-}t/2}\hat a_{n}\hat b^\dagger\right)\right\},
    \label{eq:I}\\
    &\hat{\mathcal{J}}(t):=\lambda_{0}
    \frac{\omega_{a,0}^3}{\omega_{a,1}^2\omega_b}
    \left(F^*(t)\hat b^\dagger+F(t)\hat b\right),\quad
    F(t):=i\,\frac{\omega_{a,1}^2+e^{i\omega_bt}\left\{-\omega_{a,1}^2+i\omega_{a,1}\omega_b\sin[\omega_{a,1} t]+(1-\cos[\omega_{a,1} t])\omega_b^2\right\}}{\omega_{1,+}\,\omega_{1,-}},
    \label{eq:J}
\end{align}
where  $\omega_{n,\pm}:=\omega_{a,n}\pm\omega_b$. 
$\hat{\mathcal{I}}_{n}$ denotes the direct gravitational interaction between rod A and rod B, while $\hat{\mathcal{J}}$ represents the effective coupling between rod B and the photon in cavity 1.
In addition, these operators contain an inverse of $\omega_{n,-}$, which indicates a resonance effect occurring in the limit of $\omega_{a,n}\to\omega_b$. We will see how the resonance appears in the photon interference visibility in the section \ref{sec:resonance}.

In passing, we note that the free evolution of the initial coherent state of rod A leads to a squeezed coherent state. When the photon is not in cavity 1 ($n=0$), the initial state $|\alpha\rangle_a$ is evolved by the free Hamiltonian of $\hat{a}_0$ and $\hat{a}_0^\dagger$ in Eq.~\eqref{eq:Hamiltonian a} into another coherent state $|\alpha e^{-i\omega_{a,0}t}\rangle_a$.
However, when the photon is in cavity 1, not only the optomechanical coupling is involved, but also the Hamiotonian is composed of $\hat{a}_1 $ and $\hat{a}_1^\dagger$, which are associated with the different frequency $\omega_{a,1}$. In Appendix.~\ref{apdx:time_evolution}, we show that the time-evolved state of rod A becomes a squeezed coherent state. 
However, our main result can be understood without being familiar with these lengthy calculations and intricate states.

\section{The calculation of the visibility
}\label{sec:visibility}

Based on the time-evolved state in Eq.~\eqref{eq:time_evolution}, we calculate the interference visibility of the photon in the cavities, which is defined with the absolute value of the interference term, as
\begin{align}
    \mathcal{V}_c(t)
    &:=2\left|\mathrm{Tr}\left[\,{}_{c1}\langle 0|_{c2}\langle 1|\psi(t)\rangle\langle\psi(t)|1\rangle_{c1}|0\rangle_{c2}\right]\right|,
    \notag\\
    &= \left|
    _a\langle\alpha|_b\langle\beta|e^{i\left(\hat H_{a,0}+\hat H_b\right)t/\hbar}
    \left\{1-2i g\,\left(\hat{\mathcal{I}}_0^\dagger(t)-\hat{\mathcal{I}}_1(t)-\hat{\mathcal{J}}(t)\right)\right\}
    e^{-i\left(\hat H_{a,1}+\hat H_b\right)t/\hbar}|\alpha\rangle_a|\beta\rangle_b
    +\mathcal{O}(g^2)\right|,
    \notag\\
    &= \mathcal{V}_c^{(0)}(t)
    \left(1
    +2g ~\mathrm{Im}\left[
    {}_0\langle\hat{\mathcal{I}}_0^\dagger(t)\rangle_1-{}_0\langle\hat{\mathcal{I}}_1(t)\rangle_1\right]
    \right)
    +\mathcal{O}(g^2),
    \label{eq:visibility}
\end{align}
where $\mathcal{V}_c^{(0)}$ is the result without the gravitational coupling, and
${}_0\langle\cdots\rangle_1$ is not the expectation value 
but an off-diagonal element of the photon state.
\begin{align}
    \mathcal{V}_c^{(0)}(t)
    =\left|_a\langle\alpha|e^{i\hat H_{a,0}t/\hbar}\,e^{-i\hat H_{a,1}t/\hbar}|\alpha\rangle_a\right|,
    \qquad
    {}_0\langle\cdots\rangle_1
    =\frac{_a\langle\alpha|_b\langle\beta|e^{i\left(\hat H_{a,0}+\hat H_b\right)t/\hbar}\,\cdots\,e^{-i\left(\hat H_{a,1}+\hat H_b\right)t/\hbar}|\alpha\rangle_a|\beta\rangle_b}
    {_a\langle\alpha|e^{i\hat H_{a,0}t/\hbar}\,e^{-i\hat H_{a,1}t/\hbar}|\alpha\rangle_a}.
    \label{eq:visibility result}
\end{align}
Their full expressions can be found in Appendix.~\ref{apdx:visibility}.

It is interesting to note that the contribution from $\hat{\mathcal{J}}(t)$ does not appear at $\mathcal{O}(g)$ in Eq.~\eqref{eq:visibility},
since it is a Hermitian operator and appears as $\mathrm{Im}[\langle\hat{\mathcal{J}}(t)\rangle]$, where 
$\langle\hat{\mathcal{J}}(t)\rangle= {}_b\langle\beta|e^{i\hat H_bt/\hbar}\,\hat{\mathcal{J}}(t)\,e^{-i\hat H_b t/\hbar}|\beta\rangle_b$. 
In contrast, the contribution from $\hat{\mathcal{I}}_n(t)$ survives, because the difference in the frequency of rod A, $\omega_{a,0}\neq\omega_{a,1}$, originating in the second-order contribution of $\theta_a$ distinguishes $\hat{\mathcal{I}}_0(t)$ and $\hat{\mathcal{I}}_1(t)$ and prevents their cancellation. Hence, we gain the $\mathcal{O}(g)$ contribution to the visibility.
In the previous works \cite{Balushi2018}, however, this frequency difference 
was not appreciated.
In that case, $\hat{\mathcal{I}}_1(t)$ is replaced by $\hat{\mathcal{I}}_0(t)$ and they were canceled in Eq.~\eqref{eq:visibility}. Then, the leading contribution from gravity to the visibility would be the second order of $g$, 
\begin{align}
    \mathcal{V}_c(t)
    \approx \mathcal{V}_c^{(0)}(t)
    \left(1+4 g^2 \left|\left\langle\mathcal{J}(t)\right\rangle\right|^2\right),
    \qquad (\omega_{a,0}=\omega_{a,1};\ {\rm without\ the\ higher\ order\ correction\ of\ }\theta_a).
    \label{eq:visibility_Balushi}
\end{align}
Therefore, it is possible to make a remarkable signal amplification of 
the gravitational quantum effect by considering the higher-order contribution of $\theta_a$ in our setup.

By assuming $\beta$ is a real number for simplicity, the explicit form of visibility is given by
\begin{align}
    \mathcal{V}_c(t)
    &\approx\mathcal{V}_c^{(0)}(t)\left[1+
    2g\,\omega_{a,0}\beta\left\{
    \left(\frac{\sin[\omega_{0,+}t/2]}{\omega_{0,+}}
    +\frac{\sin[\omega_{0,-}t/2]}{\omega_{0,-}}\right)
    C_{0}
    +\left(\frac{\sin[\omega_{1,+}t/2]}{\omega_{1,+}}
    +\frac{\sin[\omega_{1,-}t/2]}{\omega_{1,-}}
    \right)C_{1}
    \right\}\right],
    \label{eq:visibility2}
\end{align}
%
where we introduced $\omega_{n,-}:= \omega_{a,n}-\omega_b$.
The coefficient of each term is given by
\begin{align}
    &C_{0}=\sqrt{\frac{\omega_{a,0}}{\omega_{a,1}}}\cos\left[\frac{\omega_{a,0}t}{2}\right]
    \mathrm{Im}\left[{}_0\langle \hat{a}_1\rangle_1+{}_0\langle \hat{a}_1^\dagger\rangle_1\right]
    +\sqrt{\frac{\omega_{a,1}}{\omega_{a,0}}}\sin\left[\frac{\omega_{a,0}t}{2}\right]
    \mathrm{Re}\left[{}_0\langle \hat{a}_1\rangle_1-{}_0\langle \hat{a}_1^\dagger\rangle_1\right],\notag\\
    &C_{1}=-\sqrt{\frac{\omega_{a,1}}{\omega_{a,0}}}\left(
    \cos\left[\frac{\omega_{a,1}t}{2}\right]
    \mathrm{Im}\left[{}_0\langle \hat{a}_1\rangle_1+{}_0\langle \hat{a}_1^\dagger\rangle_1\right]
    +
    \sin\left[\frac{\omega_{a,1}t}{2}\right]
    \mathrm{Re}\left[{}_0\langle \hat{a}_1\rangle_1-{}_0\langle \hat{a}_1^\dagger\rangle_1\right]
    \right),
\end{align}
where $_0\langle\cdots\rangle_1$ was defined in
Eq.~\eqref{eq:visibility result}. 
The full expressions for $C_{n}$ without the assumption of a real $\beta$ can be found in Appendix.~\ref{apdx:visibility}.
Note that the dependence on the initial coherent state $\alpha$ of rod A and the optomechanical coupling constant $\lambda_n$ are encoded in $C_{n}$.
In the limit of $\omega_{a,1}\to \omega_{a,0}$, $C_1$ becomes $-C_0$, their coefficients become identical and the linear term of $g$ in Eq.~\eqref{eq:visibility2} vanishes.

In Eq.\eqref{eq:visibility2}, we can see that the first-order gravity-induced contribution to the visibility is proportional to $\beta$. 
These terms physically represent some supersposed states in which the oscillation of rod B gravitationally affects rod A in distinct ways depending on the frequency $\omega_{a,n}$.
The factors $\sin[\omega_{n,\pm}t/2]/\omega_{n,\pm}$ indicate that how much rod B changes the motion of rod A depends on their frequency matching between $\omega_b$ and $\omega_{a,n}$.
In particular, if they are very close $\omega_b\approx \omega_{a,n}$,
a resonance phenomenon takes place and the term with $\omega_{n,-}:= \omega_{a,n}-\omega_b$ is significantly amplified, as we will see in Sec.~\ref{sec:resonance}.

In the studies by Carney et al. \cite{Carney2021a,Carney2022}, they also investigated the appearance of the quantum gravity signals linearly dependent on the gravitational coupling constant $g$, within the context of a hybrid system comprising an oscillator and a trapped atom. They consider an initial state in which the hybrid systems are already entangled through a quantum interaction other than gravity. While their approach differs from our idea, it should be noted that they also achieved a $\mathcal{O}[g]$ contribution in the visibility of the atom by preparing the initially entangled state.

\section{The $\mathcal{O}(g)$ contribution to the visibility}\label{sec:numerical results}
In this section, we will present some numerical results demonstrating the visibility~\eqref{eq:visibility result} amplified by the new $\mathcal{O}(g)$ contribution. 
Note that we avoid the resonance in this section to separately study the two different amplification effects, and we will explore it in the next section.
We set the mirror masses of both rods $m=M=10^{-13}\,\mathrm{[kg]}$, the vertical interval between the two rods $h=2\times 10^{-6}\,\mathrm{[m]}$, the original frequency of rod A $\Omega_a=3\times 10^3\,\mathrm{[Hz]}$, the original frequency of rod B $\Omega_b=0.84\times \Omega_a\,\mathrm{[Hz]}$, the initial coherent parameter of the rods $\alpha=\beta=1$, the photon wave frequency $\omega_c=450\times 10^{12} \,\mathrm{[Hz]}$ and the original cavity length $\ell=0.01\,\mathrm{[m]}$. Using these parameters, the dimensionless parameters contained in the visibility are computed as
\begin{align}
    &\lambda_0=4.5
    \left(\frac{m}{10^{-13}\,\mathrm{[kg]}}\right)^{-1/2}
    \left(\frac{\Omega_a}{3\times10^3 \,\mathrm{[Hz]}}\right)^{-3/2}
    \left(\frac{\omega_c}{450\times 10^{12} \,\mathrm{[Hz]}}\right)
    \left(\frac{\ell}{0.01\,\mathrm{[m]}}\right)^{-1},\notag\\
    &\frac{\omega_b}{\omega_{a,0}}=1.9
    \left(\frac{\Omega_b}{3\times10^3 \,\mathrm{[Hz]}}\right)
    \left(\frac{\Omega_a}{3\times10^3 \,\mathrm{[Hz]}}\right)^{-1},
    \label{eq:parameters1}
\end{align}
where we set $\omega_b$ not very close to $\omega_{a,1}$ to avoid the resonance.
The following two parameters are especially important.
\begin{align}
    &1-\frac{\omega_{a,0}}{\omega_{a,1}}=2.8\times 10^{-10}
    \left(\frac{\lambda_0}{4.5}\right)^2
    \left(\frac{\Omega_a}{3\times10^3 \,\mathrm{[Hz]}}\right)
    \left(\frac{\omega_c}{450\times 10^{12} \,\mathrm{[Hz]}}\right)^{-1},\notag\\
    &g=5.1\times 10^{-14}
    \left(\frac{m}{10^{-13}\,\mathrm{[kg]}}\right)^{1/2}
    \left(\frac{M}{10^{-13}\,\mathrm{[kg]}}\right)^{1/2}
    \left(\frac{\Omega_a}{3\times10^3 \,\mathrm{[Hz]}}\right)^{-3/2}
    \left(\frac{h}{2\times 10^{-6}\,\mathrm{[m]}}\right)^{-3}.
    \label{eq:parameters2}
\end{align}
We see that $\omega_{a,0}/\omega_{a,1}-1$, which originates from the $\mathcal{O}[\theta_a^2]$ contribution in Eq.~\eqref{eq:thetam_expansion}, is extremely small, although the gravitational coupling parameter $g$ is even smaller.

\begin{figure}[t]
    {\centering
    \includegraphics[width=0.4\linewidth,clip]{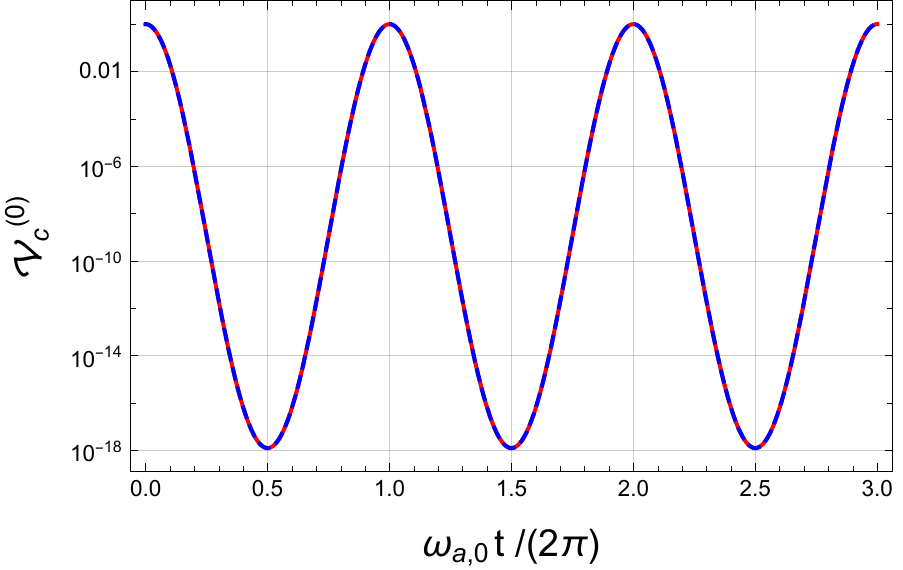}
    \hspace{5mm}
    \includegraphics[width=0.55\linewidth,clip]{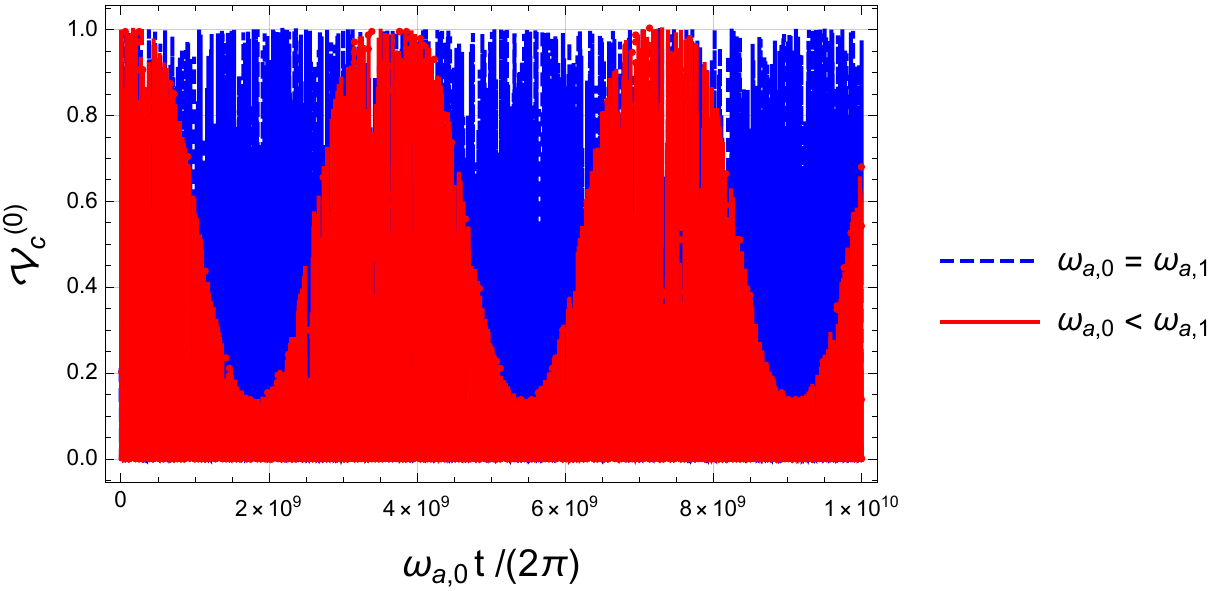}
    }
    \caption{
    The time dependence of the visibility without the gravitational contribution, $\mathcal{V}_c^{(0)}(t)$ given in Eq.~\eqref{eq:visibility result}. 
    The red line denotes our result that takes into account the higher order contribution $\mathcal{O}(\theta_a^2)$ and appreciates the frequency difference $\omega_{a,0}<\omega_{a,1}$, while the blue dashed line denotes 
    the previous result that neglects the higher order correction. The parameters are set as in Eqs.~\eqref{eq:parameters1} and \eqref{eq:parameters2}, except for $\omega_{a,1}=\omega_{a,0}$ for the blue dashed line.
    The left panel shows a log plot for an early time, and the right panel shows a linear plot for a much longer time scale. 
    As seen in the right panel, the frequency difference causes a strong dephasing 
    at the corresponding time scale, 
    $\omega_{a,0}t/(2\pi)\approx 
    1.8(2N+1)\times 10^9$, only in our result.
    }
    \label{fig:Visibility_g0}
\end{figure}
Let us first study the case without the gravitational coupling $g=0$, namely the 0-th order results.
Fig.~\ref{fig:Visibility_g0} shows the time dependence of the visibility $\mathcal{V}_c^{(0)}$. The left panel shows the behavior in the early time and the right panel shows for the longer time period. The red lines represent the result of our calculation that takes the $\mathcal{O}[\theta_a^2]$ contribution into account, leading to $\omega_{a,0}< \omega_{a,1}$. 
The blue dashed lines ignore the correction and adopt $\omega_{a,0}= \omega_{a,1}$ in the same way as the previous works \cite{Balushi2018}. 
As seen in the left panel, the visibility decoheres and recoheres due to the optomechanical coupling between the photon and the rod A systems.
No visible difference between the two cases is observed for the early time. 
However, we see a clear difference in the photon visibility in the right panel of Fig.~\ref{fig:Visibility_g0}, 
which comes from the frequency difference in $\omega_{a,n}$ even without the gravitational coupling. 
This strong dephasing at around $\omega_{a,0}t/(2\pi)\approx 2\times 10^9$ is
caused by the fact that
the two states of rod A with and without the photon, namely $e^{-i\hat H_{a,0}t/\hbar}|\alpha\rangle_a$ and $e^{-i\hat H_{a,1}t/\hbar}|\alpha\rangle_a$, oscillate  
for the different time scales $1/\omega_{a,0}$ and $1/\omega_{a,1}$, respectively.
These two states become nearly orthogonal for every period of time
when their phase difference accumulates to $(2N+1)\pi$,
\begin{align}
    (\omega_{a,1}-\omega_{a,0})t = (2N+1)\pi
    \quad \Longrightarrow\quad
    \frac{\omega_{a,0}t}{2\pi} = 
    \frac{N+1/2}{\omega_{a,1}/\omega_{a,0}-1}
    \approx 1.8(2N+1)\times 10^9,
    \label{eq:wa time scale}
\end{align}
where $N=0,1,2,...$ is integer.
This explains why the recoherence of the red line is repeatedly suppressed in the right panel of Fig.~\ref{fig:Visibility_g0}.

\begin{figure}[t]
    \centering
    \includegraphics[width=0.45\linewidth]{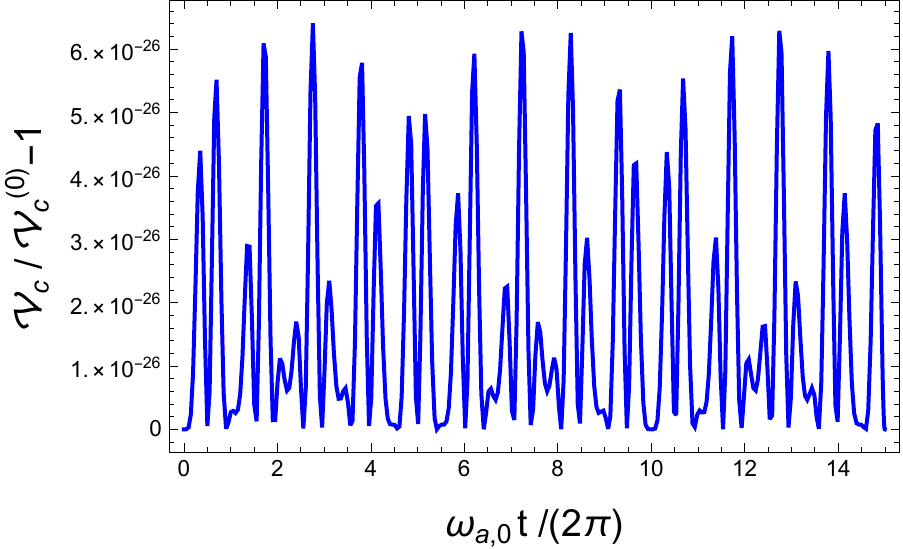}
    \hspace{5mm}
    \includegraphics[width=0.45\linewidth]{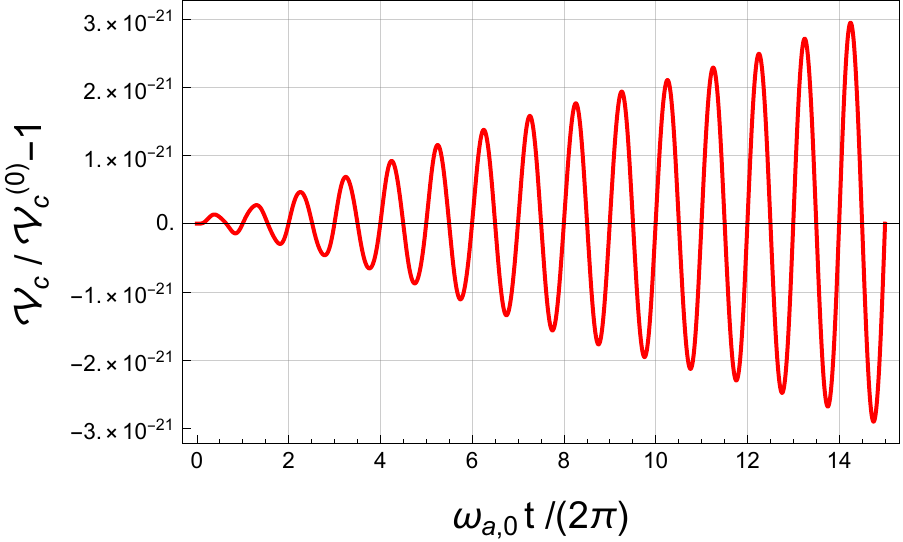}
    \caption{
    The time dependence of the gravitational contribution to the visibility, $\mathcal{V}_c/\mathcal{V}_c^{(0)}-1$. The parameters are set as in Eqs.~\eqref{eq:parameters1} and \eqref{eq:parameters2}. The left panel shows the result in Eq.~\eqref{eq:visibility_Balushi} when we neglect the higher order contribution of $\mathcal{O}[\theta_a^2]$, or namely $\omega_{a,0}=\omega_{a,1}$. We see a periodic 
    recoherence 
    whose amplitude is order estimated as $\mathcal{O}[4g^2\lambda_0^2]$.
    The right panel displays the result in Eq.~\eqref{eq:visibility2},  
    which takes into account the $\mathcal{O}[\theta_a^2]$ contribution and  $\omega_{a,0}<\omega_{a,1}$.  
    Its amplitude $\mathcal{O}\left[4g(\alpha+\lambda_0)(1-\omega_{a,0}/\omega_{a,1})\right]\times\omega_{a,0} t$  
    in this plot is larger than one in the left panel only by a factor of $\sim 10^4$.
    However, we will see  
    a much greater growth of the visibility change at a sufficiently longer time scale in Fig.~\ref{fig:Visibility_gdifference_longperiod}.
    }
    \label{fig:Visibility_gdifference_shortperiod}
\end{figure}

\begin{figure}[ht]
    \centering
    \includegraphics[width=0.45\linewidth]{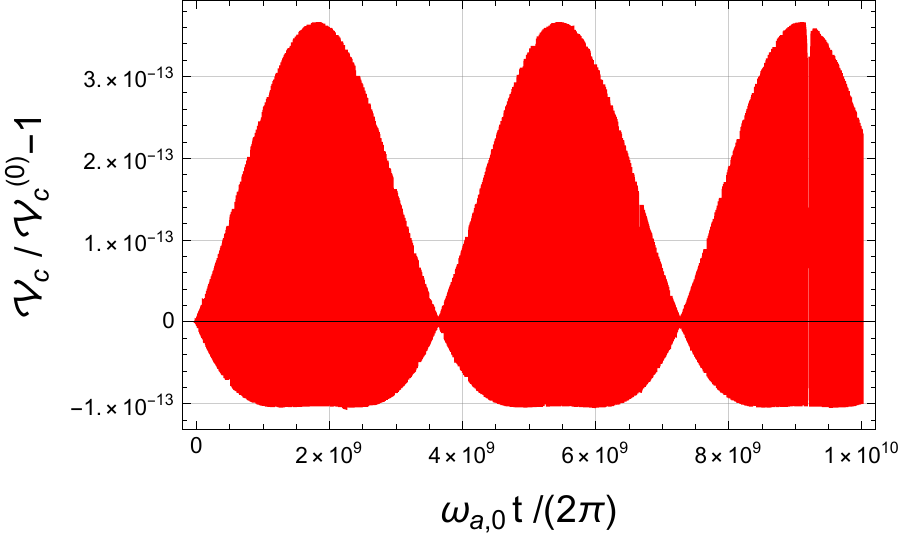}
    \caption{ 
    The gravitational contribution to the visibility given in Eq.~\eqref{eq:visibility2}
    is shown for a longer time scale. We consider the higher-order optomechanical contribution $\mathcal{O}[\theta_a^2]$, which gives $\omega_{a,0}<\omega_{a,1}$. 
    At the dephasing time derived in Eq.~\eqref{eq:wa time scale}, we observe a large amplification of the gravitational signal about $\mathcal{O}[4g(\alpha+\lambda_0)]$, denoted by the first order of the gravitational coupling $g$. In contrast, the $\omega_{a,0}=\omega_{a,1}$ result shown in the left panel of Fig.~\ref{fig:Visibility_gdifference_shortperiod} was given by the second-order of $g$, namely $\mathcal{O}[4g^2\lambda_0^2]$.
    Hence, the signal is enhanced due to 
    the reduction of the gravitational coupling order from $\mathcal{O}[g^2]$ into $\mathcal{O}[g]$ by taking the higher-order contribution $\mathcal{O}[\theta_a^2]$ into account.}
    \label{fig:Visibility_gdifference_longperiod}
\end{figure}

In Fig.~\ref{fig:Visibility_gdifference_shortperiod} and Fig.~\ref{fig:Visibility_gdifference_longperiod},
we present the gravitational contribution to the visibility
as the relative correction from the no gravity cases seen above, $\mathcal{V}_c(t)/\mathcal{V}_c(t)^{(0)}-1$. 
The parameters are the same as Eqs.~\eqref{eq:parameters1} and \eqref{eq:parameters2} again.
The left panel of Fig.~\ref{fig:Visibility_gdifference_shortperiod} depicts the result for the $\omega_{a,0}=\omega_{a,1}$ case as in Ref.\cite{Balushi2018}.
We see a periodic motion of the visibility correction from gravity. Its amplitude is roughly estimated from Eq.~\eqref{eq:visibility_Balushi} as $4g^2 \left|\left\langle\mathcal{J}(t)\right\rangle\right|^2\approx\mathcal{O}[4g^2\lambda_0^2]\approx 9.4\times10^{-26}$.
The right panel of Fig.~\ref{fig:Visibility_gdifference_shortperiod} shows the result in the $\omega_{a,0}<\omega_{a,1}$ case respecting the second-order contribution of $\theta_a$. 
We can see that the visibility repeats decoherence and recoherence which amplitude is linear growing. 
We can derive the growth rate of the oscillation from Eq.~\eqref{eq:visibility2}. If we replace periodic functions contained in $C_0,~C_1$ to $1$, we get order estimation of these functions as $C_0\approx -C_1\approx \mathcal{O}\left[2(\alpha+\lambda_0)\right]$, which indicates that the initial coherent state of rod A is displaced by $\lambda_0$ due to the photon pressure. By substituting these estimations into Eq.~\eqref{eq:visibility2} and considering a leading term of $1-\omega_{a,0}/\omega_{a,1}$, we obtain
\begin{align}
    \mathcal{V}_c(t)/\mathcal{V}_c(t)^{(0)}-1
    \approx
    \mathcal{O}\left[
    4 g (\alpha+\lambda_0)\left(1-\frac{\omega_{a,0}}{\omega_{a,1}}\right)
    \right]\times
    \omega_{a,0}t
    \approx1.3\times 10^{-21}~\frac{\omega_{a,0}t}{2\pi}\,.
\end{align}
This estimation holds in a short time scale satisfying $t\ll (\omega_{a,1}-\omega_{a,0})^{-1}$ as in the right panel of Fig.~\ref{fig:Visibility_gdifference_shortperiod}. This is about $\mathcal{O}\left[(1-\omega_{a,0}/\omega_{a,1})(\alpha+\lambda_0)/(g^2\lambda_0^2)\right]\omega_{a,0}t\approx1.4\times 10^4~\omega_{a,0}t/(2\pi)$ larger compared to the $\omega_{a,0}=\omega_{a,1}$ case in the left panel.
In Fig.~\ref{fig:Visibility_gdifference_longperiod}, we present the gravitational contribution to the visibility for a longer time period in our case of $\omega_{a,0}<\omega_{a,1}$. Again, we observe the periodic dephasing at $\omega_{a,0}t/(2\pi)
\approx 1.8(2N+1)\times 10^9$ as explained in Eq.~\eqref{eq:wa time scale}.
The amplitude reaches an order of $10^{-13}$ at that time. In contrast, in the $\omega_{a,0}=\omega_{a,1}$ case, the amplitude does not exceed $\sim 10^{-23}$ even in the longer time scale. 

The significant amplification of the longer time period in Fig.~\ref{fig:Visibility_gdifference_longperiod} arises because the visibility is given by the first order of $g$ in our case, whereas it appears from the second order of $g$ if we disregard the second-order contribution of $\theta_a$. After a sufficient amount of time has passed, the terms inside the bracket $\{\cdots\}$
in Eq.~\eqref{eq:visibility2} become $\mathcal{O}\left[2(\alpha+\lambda_0)
\right]$, and the gravitational shift of the visibility extends to $\mathcal{O}\left[4 g(\alpha+\lambda_0)\right]$,
which is on the order of $7.5\times 10^{-13}$ and consistent with Fig.~\ref{fig:Visibility_gdifference_longperiod}.
It should be noted that we chose the value of $\omega_b$ for which the resonance is ineffective, and thus, this amplification of the visibility results only from the reduction of the order of the gravitational coupling $g$. 
In the next section, we will discuss how to further enhance the gravitational signal in visibility using the resonance effect.

\section{The resonance effect
}\label{sec:resonance}

Since the setup contains two oscillators, we expect that a resonant behavior affects the visibility if their frequencies are close enough. In this section, we discuss the case where $\omega_{a,1}$ is close to $\omega_b$, focus on the resonance term in the visibility in Eq.~\eqref{eq:visibility2}, and discuss how much the resonance effect amplifies the visibility.

We will consider the resonance for $\omega_b\approx \omega_{a,1}$. 
This physically means that the oscillating rod A resonates with rod B only when the photon enters cavity 1. Since the visibility captures the state difference between the photon within cavity 1 and cavity 2, the resonance effect is supposed to affect  
the visibility significantly.
However, remember that $\omega_{a,1}$ and $\omega_{a,0}$ are very close as seen in Eqs.~\eqref{eq:parameters2}. 
Therefore, when we suppose to set $\omega_b$ to be close to $\omega_{a,1}$, 
$\omega_b$ is inevitably close to $\omega_{a,0}$ as well.
If $\omega_{a,1}$ is closer to $\omega_{b}$ much more than $\omega_{a,0}$, the system has a exclusive resonance only between $\omega_b$ and $\omega_{a,1}$. Then, the strength of the resonance effect 
is controlled by their frequency difference.
We introduce such a frequency matching parameter as
\begin{align}
\epsilon:=\frac{\omega_{1,-}}{\omega_{a,1}}=1-\frac{\omega_b}{\omega_{a,1}}\,.    
\end{align}
In contrast, if $\omega_{a,1}$ is much closer to $\omega_{a,0}$ than $\omega_{b}$, that is $\omega_{b}$ is close to both of $\omega_{a,1}$ and $\omega_{a,0}$, the resonance takes place in both superposed states simultaneously. Then the resonant contribution from the gravitational coupling to the visibility is suppressed, because this effect does not distinguish the two superposed states labeled by $n=0$ and $n=1$.
To determine which of the above two cases happens, we compare $\epsilon$ to $1-\omega_{a,0}/\omega_{a,1}$.
For $\epsilon\ll 1-\omega_{a,0}/\omega_{a,1}$,
the exclusive resonance occurs, while the simultaneous resonance takes place for $\epsilon\gg 1-\omega_{a,0}/\omega_{a,1}$.
We will confirm this physical argument by analytic and numerical investigations below.

Assuming $\alpha=0$ and $\beta\in \mathbb{R}$ to simplify the expression, 
Eq,~\eqref{eq:visibility2} reduces to
\begin{align}
    \mathcal{V}_C(t)\approx \mathcal{V}_C^{(0)}(t)
    \left\{1-
    2 g \lambda_0 \beta\omega_{a,0}
    \left(\frac{\sin[\omega_{1,-}t/2]}{\omega_{1,-}}-\frac{\sin[\omega_{0,-}t/2]}{\omega_{0,-}}\right)
    \left(
    \sin\left[\frac{\omega_{1,+}t}{2}\right]+\sin\left[\frac{\omega_{1,-}t}{2}\right]
    \right)\right\}
    \label{eq:visibility_resonance}
\end{align}
The second term denotes the gravitational contribution to the visibility in $\mathcal{O}[g]$ and can exhibit the resonance. 
If we make a measurement at some time around $t\approx 1/\omega_{n,-}$, the resonance effect would be significant. 
Particularly, at around the time $t\approx \pi/\omega_{1,-}$, we obtain
\begin{align}
    \frac{\mathcal{V}_C(t)}{\mathcal{V}_C^{(0)}(t)}
    &\approx 
    1-
    \left(1+
    \sin\left[\frac{\omega_{1,+}t}{2}\right]
    \right)
    \times
    \left\{
    \begin{array}{ll}
        2 g \lambda_0 \beta/\epsilon \quad 
        & (\epsilon \ll 1-\frac{\omega_{a,0}}{\omega_{a,1}}\text{ : Exclusive resonance}) \\
        2 g \lambda_0 \beta\left(1-\frac{\omega_{a,0}}{\omega_{a,1}}\right)/\epsilon^2 \quad 
        & (\epsilon\gg 1-\frac{\omega_{a,0}}{\omega_{a,1}}\text{ : Simultaneous resonance})
    \end{array}
    \right.\,,
    \label{eq:visibility_resonance2}
\end{align}
where we used $\omega_{0,-}/\omega_{a,0} = \epsilon +(1-\omega_{a,0}/\omega_{a,1})+\mathcal{O}\left[(1-\omega_{a,0}/\omega_{a,1})^2\right]$ to obtain the expression on the lower case.
The upper case indicates the exclusive resonance
and the lower case corresponds to the simultaneous resonance.
Compared to the upper case, the lower case
is suppressed by a factor of $(1-\omega_{a,0}/\omega_{a,1})/\epsilon\ll 1$.

\begin{figure}[t]
    \centering
    \includegraphics[width=0.65\linewidth]{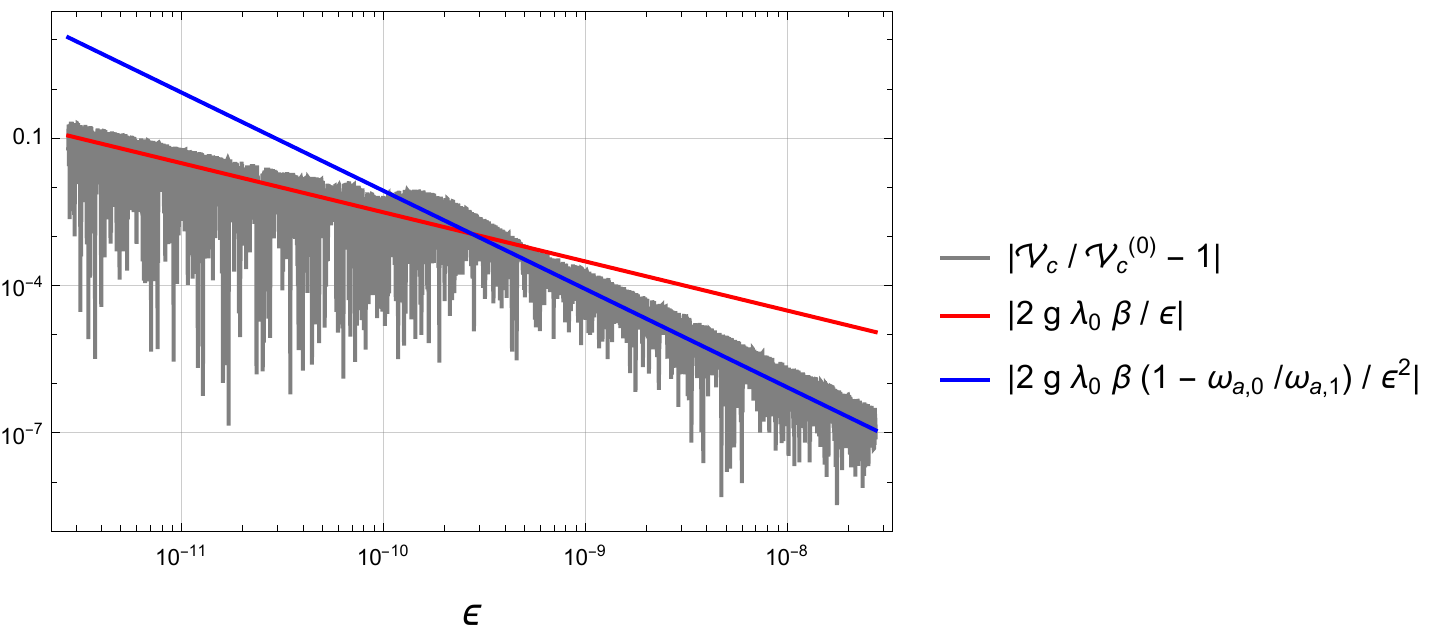}
    \caption{Resonance behavior of the gravitational correction to the visibility against the frequency matching parameter $\epsilon:=1-\omega_b/\omega_{a,1}$. The gray line denotes the relative contribution from gravity to the visibility, $|\mathcal{V}_c/\mathcal{V}_c^{(0)}-1|$ at the time of $t=\pi/\omega_{1,-}$.
    The red and blue lines represent the analytic estimate of the resonance effect given in Eq.~\eqref{eq:visibility_resonance2} for $1-\omega_{a,0}/\omega_{a,1}\gg\epsilon$ and $1-\omega_{a,0}/\omega_{a,1}\ll\epsilon$,  respectively.
    The red and blue lines show good agreements with the numerical result on the left and right region of $\epsilon=1-\omega_{a,0}/\omega_{a,1}=2.8\times 10^{-10}$ respectively as expected.
    The parameters are set as 
    $\lambda_0=4.5,~
    1-\omega_{a,0}/\omega_{a,1}=2.8\times 10^{-10},~
    g=5.1\times 10^{-14},~
    \alpha=0,~\beta=1$.}
    \label{fig:Visibility_resonance}
\end{figure}

Fig.~\ref{fig:Visibility_resonance} shows the resonance behavior of the visibility for the varying frequency matching parameter $\epsilon$. A gray line denotes the absolute value of the relative modification of the visibility due to gravity $|\mathcal{V}_c/\mathcal{V}_c^{(0)}-1|$
at the observation time $t=\pi/\omega_{1,-}$. Red and blue lines represent the absolute value of the  
last factor in Eq.~\eqref{eq:visibility_resonance2} for $\epsilon\ll 1-\omega_{a,0}/\omega_{a,1}$ and $\epsilon\gg1-\omega_{a,0}/\omega_{a,1}$, respectively, namely
$\left|2 g\lambda_0\beta/\epsilon\right|$ and $|2 g \lambda_0 \beta(1-\omega_{a,0}/\omega_{a,1})/\epsilon^2|$.
Note that we set $\alpha=0$ to justify the assumption of Eq.~\eqref{eq:visibility_resonance}. 
The other parameters are chosen as in Eqs.~\eqref{eq:parameters1} and \eqref{eq:parameters2}. 
The red and blue lines agree well with the numerical calculation 
in the corresponding parameter regions.
As we expected, the resonance enhancement is characterized by the inverse of $\epsilon$ on the left side, while the double inverse of $\epsilon$ on the right side. We also observe that the transition takes place when the $\omega_{a,0}$ resonance becomes comparable with the $\omega_{a,1}$ resonance at $\epsilon\approx (1-\omega_{a,0}/\omega_{a,1})=2.8\times 10^{-10}$. 

\begin{figure}[htbp]
    \centering
    \includegraphics[width=0.45\linewidth]{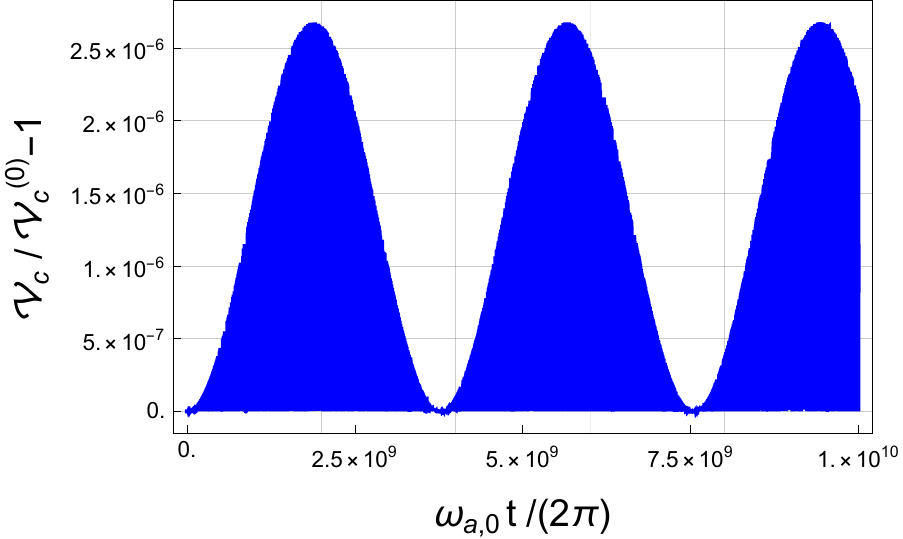}
    \hspace{5mm}
    \includegraphics[width=0.45\linewidth]{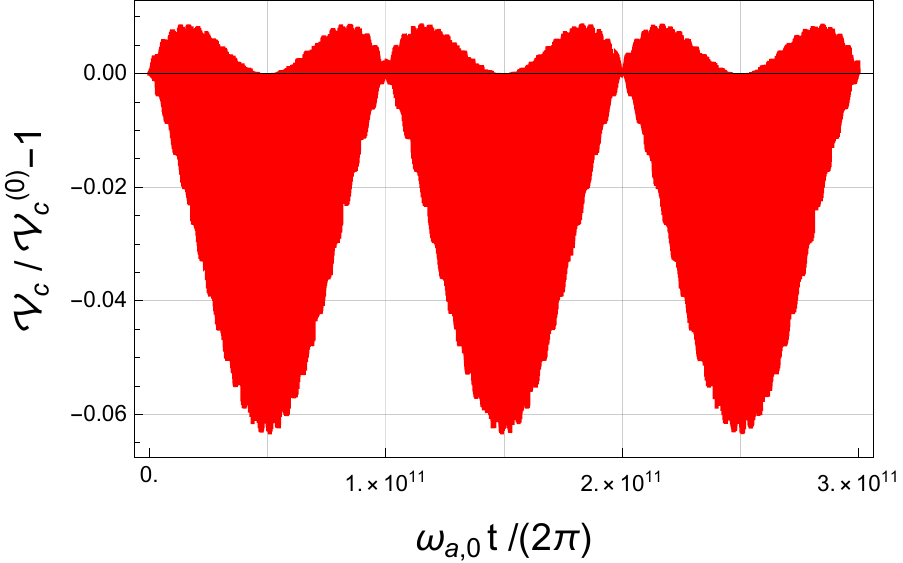}
    \caption{
    The gravitational contribution to the visibility $\mathcal{V}_c/\mathcal{V}_c^{(0)}-1$ enhanced by the resonance effect for the long time scale. We set parameters as $\epsilon:=1-\omega_b/\omega_{a,1}=10^{-11}$ to induce the resonance and $\alpha=0$ for simplicity. Otherwise, we choose the same parameters as given in Fig.~\ref{fig:Visibility_gdifference_shortperiod} and \ref{fig:Visibility_gdifference_longperiod}.  The left panel shows the case when we ignored the higher order contribution $\mathcal{O}[\theta_a^2]$, i.e. $\omega_{a,0}=\omega_{a,1}$, which is evaluated using Eq.~\eqref{eq:visibility_Balushi}. This corresponds to the resonant version of the left panel in Fig.~\ref{fig:Visibility_gdifference_shortperiod}, and we see a resonance enhancement of $\mathcal{O}[(\omega_{0,-}/\omega_{a,0})^{-2}]\sim 10^{20}$ 
    compared to the result given in the previous section. The right panel shows the result when we take $\mathcal{O}[\theta_a^2]$ into account, i.e. $\omega_{a,0}<\omega_{a,1}$, which is given in Eq.~\eqref{eq:visibility2}. This result is the resonant version of Fig.~\ref{fig:Visibility_gdifference_longperiod}, and is about $\mathcal{O}[\epsilon^{-1}]\sim 10^{11}$ times larger than the result in Fig.~\ref{fig:Visibility_gdifference_longperiod}.
    }
    \label{fig:Visibility_gdifference_resonance}
\end{figure}

In Fig.~\ref{fig:Visibility_gdifference_resonance}, we present the gravitational contribution to the visibility with parameters yielding the resonance effect. 
We take $\omega_b/\omega_{a,0}\approx 1+2.7\times 10^{-10}$, which corresponds to $\epsilon=10^{-11}$; 
This indicates the exclusive resonance of $\omega_{a,1}$ and $\omega_b$ which we find in the left region in Fig.~\ref{fig:Visibility_resonance}.
Otherwise, we adopt the parameters in Eq.~\eqref{eq:parameters1} and \eqref{eq:parameters2}. The left and the right panels show the $\omega_{a,0}=\omega_{a,1}$ case and the $\omega_{a,0}<\omega_{a,1}$ case, respectively. Comparing with Fig.~\ref{fig:Visibility_gdifference_shortperiod} and \ref{fig:Visibility_gdifference_longperiod}, we see the significant enhancement of the amplitude in both panels arising from the resonance effect. Note that the resonance also occurs even if we ignore the second order of $\theta_a$ as seen in the left panel of Fig.~\ref{fig:Visibility_gdifference_resonance}. This is because the visibility correction in Eq.~\eqref{eq:visibility_Balushi} is given by $\hat{\mathcal{J}}$, which also contains a term inversely proportional to $\omega_{1,-}=\omega_{0,-}$ (see Eq.~\eqref{eq:J}). This leads to a periodic enhancement in the visibility change of $\mathcal{O}\left[4g^2\lambda_0^2/(\omega_{0,-}/\omega_{a,0})^2\right]\approx 1.3\times10^{-6}$ with a time scale $\omega_{a,0} t\approx (\omega_{0,-}/\omega_{a,0})^{-1}\approx 3.8\times 10^9$. 
In the right panel of Fig.~\ref{fig:Visibility_gdifference_resonance}, we see an even larger amplitude of the visibility change,   
which reaches the percent level.
The resonance effect amplifies the result of Fig.~\ref{fig:Visibility_gdifference_longperiod},
whose amplitude was $\mathcal{O}[4g(\alpha+\lambda_0)]$,
by the factor of $\mathcal{O}[1/\epsilon]$
and achieves the amplitude of
$\mathcal{O}\left[4g(\alpha+\lambda_0)/\epsilon\right]\approx 7.5\times 10^{-2}$ periodically with a time scale $\omega_{a,0} t\approx 1/\epsilon=10^{11}$.

To see the resonant amplification of $1/\epsilon$ as in the right panel of Fig.~\ref{fig:Visibility_gdifference_resonance}, the system is required to maintain its quantum coherence for about 3 years, $t \approx  1/(\omega_{a,0}\epsilon) \approx 3.3 \times 10^7 \text[s]$, with our parameter choice. However, this is technically difficult to achieve at present due to the environmental decoherence of the quantum system. Also, it is challenging to tune two frequencies to be sufficiently close with high accuracy of $\epsilon=10^{-11}$. These difficulties indicate that there is a lower bound of $\epsilon$ in the realistic situation. Let us suppose that $\epsilon$ is fixed at some value in the right region of Fig.~\ref{fig:Visibility_resonance} regarding as the possible lower bound in the setup, and control the parameter $1-\omega_{a,0}/\omega_{a,1}$ to obtain the best resonance enhancement. As we raise $1-\omega_{a,0}/\omega_{a,1}$, the blue plot in Fig.~\ref{fig:Visibility_resonance} shifts upward, which means that we gain more enhancement at fixed $\epsilon$. Hence, if the experimental setup is possible to achieve $\epsilon\,(>1-\omega_{a,0}/\omega_{a,1})$, we observe the signal enhancement of $2g\lambda_0\beta(1-\omega_{a,0}/\omega_{a,1})/\epsilon^2$ due to the resonance effect, which amplification is improved by setting a larger $1-\omega_{a,0}/\omega_{a,1}$.

To summarize the results of Sec.~\ref{sec:numerical results} and Sec.~\ref{sec:resonance}, we found two ways to enhance the gravitational contribution to the visibility; First, we take the second-order term in the optomechanical coupling into account, which leads to $\omega_{a,0}<\omega_{a,1}$. Then, the visibility is given by the first order of the gravitational coupling $g$ at $t\approx (\omega_{a,1}-\omega_{a,0})^{-1}$, while it appears from its second order in the previous works \cite{Balushi2018}. 
This first effect can amplify the signal by a factor of $1/(g\lambda_0)$.
Second, by adjusting the frequencies of two oscillators to be close enough $\epsilon:=1-\omega_b/\omega_{a,1}\ll1$, we gain a resonance effect depending on the parameter regions.  For $\epsilon \ll 1-\omega_{a,0}/\omega_{a,1}$, rod A resonates with rod B only when the photon enters cavity 1, and the signal gains $1/\epsilon$ amplification due to this exclusive resonance.  While for $\epsilon \gg 1-\omega_{a,0}/\omega_{a,1}$, each oscillating mode of rod A resonates with rod B
, and the signal is amplified about $(1-\omega_{a,0}/\omega_{a,1})/\epsilon^2$ due to the simultaneous resonance.
Finally, the signal on the right panel in Fig.~\ref{fig:Visibility_gdifference_resonance} is $1/(g\lambda_0\epsilon)\sim 10^{24}$ times amplified compared to the original result on the left panel in Fig.~\ref{fig:Visibility_gdifference_shortperiod}.

\section{Gravity-induced entanglement}\label{sec:entanglement}

The quantum entanglement \cite{Horodecki2009} created by gravity between systems is one of the  
major targets to probe the quantum feature of gravity \cite{Marletto2017,Bose2017}. 
In this section, we adopt the entanglement negativity as a measure of quantum entanglement; Negativity of bipartite states is defined as the sum of negative eigenvalues of the partially transposed density matrix \cite{Vidal2002, Horodecki1996, Peres1996}. 
This is closely related to the maximum number of distillable Bell pairs in the system. 
Especially, the value of negativity of a state vanishes when the state is separable, and takes $1/2$ when the state is given by the Bell state.
We evaluate the negativity between rod B and the other systems which should be induced by the quantum gravitational interaction between the two rods. 

To obtain the negativity, we calculated the partially transposed total density matrix $\hat \rho^{\text{T}_\text{B}}(t)$ and expand it with respect to the small parameter $g$.
Then, we compute its eigenvalues up to the first order of $g$. The negativity between rod B and the others is given by a summation of the negative eigenvalues of $\hat \rho^{\text{T}_\text{B}}(t)$. As a result, we obtain the following expressions,
\begin{align}
    &\mathcal{N}_{B:A+c}
    =2g\sqrt{\sum_{n=0,1} {}_a\langle\alpha|
    e^{i\hat H_{a,n}t/\hbar}\hat{\mathcal{K}}^{\dagger}_n(t)
    \hat{\mathcal{K}}_n(t)e^{-i\hat H_{a,n}t/\hbar}|\alpha\rangle_a}\,,
    \label{eq:negativity}\\
    &\hat{\mathcal{K}}_n(t)=\sqrt{\frac{\omega_{a,0}^3}{\omega_{a,n}}}
    \left(
    \frac{\sin[\omega_{n,+}t/2]}{\omega_{n,+}}e^{i\omega_{n,+}t/2}\hat a_n
    +\frac{\sin[\omega_{n,-}t/2]}{\omega_{n,-}}e^{-i\omega_{n,-}t/2}\hat a_n^\dagger
    +n \lambda_0
    \left(\frac{\omega_{a,0}}{\omega_{a,1}}\right)^{3/2} 
    \frac{F(t)}{\omega_b}
    \right)\,.
    \label{eq:K}
\end{align}
Even in the limit of $\omega_{a,1}\to\omega_{a,0}$, we find a non-zero value of the negativity~\eqref{eq:negativity} in the first order of $g$, although the gravitational contribution in visibility appears only from its second order. This implies that the entanglement generation 
reflects in the gravitational correction to the visibility only in a very suppressed way, when we ignore the higher order optomechanical contribution $\mathcal{O}[\theta_a^2]$. In the meantime, the operator $\hat{\mathcal{K}}_n$ is closely related to  
$\hat{\mathcal{I}}_n$ and $\hat{\mathcal{J}}$, which are used in the calculation of the visibility and given in Eqs.~\eqref{eq:I} and \eqref{eq:J}, as $\hat{\mathcal{I}}_n+n\hat{\mathcal{J}}=
    \hat{\mathcal{K}}_n\hat{b}    +\hat{\mathcal{K}}^\dagger_n\hat{b}^\dagger$.

To explore how the negativity and the visibility are related, we simplify Eq.~\eqref{eq:negativity} under several assumptions. We focus on the situation where 
$\omega_{a,1}$ is much closer to $\omega_b$ than $\omega_{a,0}$,
and the resonance due to $\omega_{a,1} \approx \omega_b$ 
exclusively takes place.
Its condition is given by $1\gg 1-\omega_{a,0}/\omega_{a,1}\gg\epsilon$.
In addition, we assume $\alpha=0$ for simplicity. 
We also make use of the relation $\lambda_{0}^2\gg 1-\omega_{a,0}/\omega_{a,1}$, which means that the second-order contribution of $\theta_a$ is sub-dominant compared to its first order contribution. Then the negativity reduces to the following form.
\begin{align}
    \mathcal{N}_{B: A+c}
    \approx 2g\,\omega_{a,0}\lambda_0
    \left|\frac{\sin[\omega_{1,-}t/2]}{\omega_{1,-}}\right|\,.
\end{align}
Here,  
we see that the resonance effect amplifies the negativity, if we wait until $t\approx 1/\omega_{1,-}$, in the same way as the visibility.
By comparing this simplified form of negativity to the visibility in Eq.~\eqref{eq:visibility_resonance} under the assumption $1\gg 1-\omega_{a,0}/\omega_{a,1}\gg \epsilon >0$, we acquire a relationship between visibility and negativity as
\begin{align}
    \mathcal{V}_c(t)
    \approx
    \mathcal{V}^{(0)}_c(t)\left[1-
  \beta
    \,\mathcal{N}_{B: A+c}(t)\times\left\{
    \left|\sin\left[\frac{\omega_{1,-}t}{2}\right]\right|
    +
    \mathrm{sgn}\left[\sin\left[\frac{\omega_{1,-}t}{2}\right]\right]
    \sin\left[\frac{\omega_{1,+}t}{2}\right]
    \right\}\right]
    \label{eq:visibility and negativity}
\end{align}
The second term on the right-hand side  is proportional to the negativity, and it clearly indicates that the visibility of the photon system alters due to the gravity-induced entanglement between rod B and other systems. Moreover, the last term depending on $\omega_{1,+}$ is a highly oscillating mode, and the visibility behavior in a long time scale is almost determined by $|\sin[\omega_{1,-}t/2]/\omega_{1,-}|$ under the assumptions we imposed. Hence, when the resonance effect of the visibility exists, the production of the gravity-induced entanglement is also amplified due to the resonance. 
Comparing Fig.~7 with the right panel of Fig.~6, the time scale that the negativity grows is approximately the same as it that $\mathcal{V}_c/\mathcal{V}^{(0)}_c -1$ has a large negative value, that is, the visibility $\mathcal{V}_c$ degrades due to gravity. 
This means that gravity-induced entanglement can lead to the decoherence of the photon.

\begin{figure}[t]
    \centering
    \includegraphics[width=0.55\linewidth]{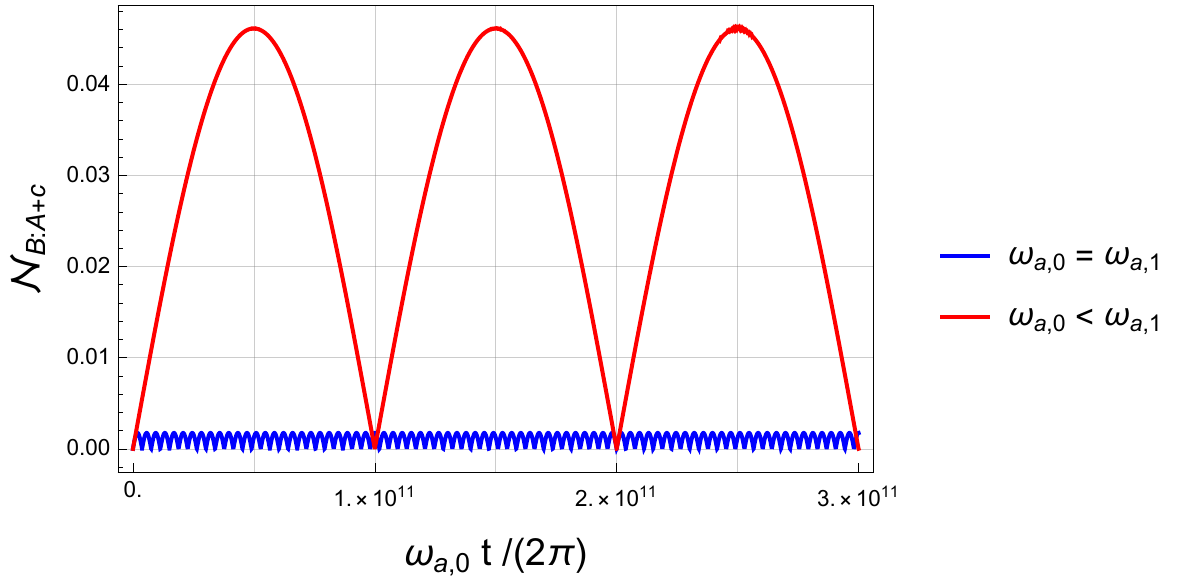}
    \caption{
    Time dependence of the quantum entanglement generated between rod B and the others (i.e. rod A and the photon) in the resonant scenario. The vertical axis denotes a measure of entanglement called negativity given in Eq.~\eqref{eq:negativity}, which value takes zero for a separable state. A red line shows the result when we consider the higher-order contribution $\mathcal{O}[\theta_a^2]$, that leads to $\omega_{a,0}<\omega_{a,1}$. A blue line shows the case when we ignored $\mathcal{O}[\theta_a^2]$, or equivalently $\omega_{a,0}=\omega_{a,1}$. The parameters are the same as in Fig.~\ref{fig:Visibility_gdifference_resonance}. The negativity behaves along with the visibility in Fig.~\ref{fig:Visibility_gdifference_resonance} as expected from their relation Eq.~\eqref{eq:visibility and negativity}. The resonance amplification can be seen in both figures. 
    }
    \label{fig:Negativity}
\end{figure}

In Fig.~\ref{fig:Negativity}, we present the time dependence of the negativity between rod B and the others in the resonant case. A red line denotes the $\omega_{a,0}<\omega_{a,1}$ case, while a blue line denotes the $\omega_{a,0}=\omega_{a,1}$ case. We take the resonance parameters which are the same as in Fig.~\ref{fig:Visibility_gdifference_resonance}; $\lambda_0=4.5,~\omega_b/\omega_{a,0}\approx1+2.7\times 10^{-10}$ (i.e. $\epsilon=10^{-11}$), $1-\omega_{a,0}/\omega_{a,1}=2.8\times 10^{-10},~g=5.1\times 10^{-14},~\alpha=0,~\beta=1$.
For $\omega_{a,0}<\omega_{a,1}$ case, we see the amplitude is enhanced by $\mathcal{O}[2g\lambda_0/\epsilon]\approx3.1\times 10^{-2}$ periodically with a time scale $1/\epsilon=10^{11}$. Also, for  $\omega_{a,0}=\omega_{a,1}$ case, the resonance enhancement is about $\mathcal{O}[2g\lambda_0(\omega_{a,0}/\omega_{0,-})]\approx1.2\times 10^{-3}$ with its time period $\omega_{a,0}/\omega_{0,-}\approx3.8\times 10^9$. 
It should be noted that the negativity is given by the first order of the gravitational coupling $g$ even for $\omega_{a,0}=\omega_{a,1}$ case as shown in Eq.~\eqref{eq:negativity}, while the visibility appears from its second order. This implies that the entanglement generation is not fully captured in the visibility when we ignore the higher order optomechanical contribution $\mathcal{O}[\theta_a^2]$.
Comparing the two cases, we find that the amplitude of $\omega_{a,0}<\omega_{a,1}$ case is about $10$ times larger than $\omega_{a,0}=\omega_{a,1}$ case, which arises from the exclusive resonance effect as in Fig.~\ref{fig:Visibility_gdifference_resonance}. Also, we find that the visibility decoheres as the negativity increases by comparing Fig.~\ref{fig:Visibility_gdifference_resonance} and \ref{fig:Negativity}, as we expected from Eq.~\eqref{eq:visibility and negativity}. Physically, this result indicates that the optomechanical system decoheres due to the entanglement generation between rod B and the photon systems.

\section{Discussion and Conclusion}\label{sec:conclusion}

Today, numerous experimental approaches are proposed to discover the quantum aspect of gravity. However, nobody has observed the quantum gravitational signal yet. 
Recently, inspired by the experimental progress in optomechanical systems, the optomechanical Cavendish experiment was proposed as a realistic way to probe the quantum nature of gravity \cite{Balushi2018}. 
Based on the previous research \cite{Balushi2018,Matsumura2020}, 
we considered an experimental setup with an optical cavity system and two mechanical rods A and B.
In the setup, a cavity photon is coupled to rod A, and two rods A and B gravitationally interacts.
We suppose to read the quantum gravity effect from the interference visibility of the photon. 
In contrast to the previous research \cite{Balushi2018,Matsumura2020}, it should be remarked that we treat up to the second order of the oscillation angle of rod A, $\theta_a$, which is considered as a higher order of optomechanical coupling between the photon and rod A systems. According to the first order of optomechanical coupling, the rod A state evolves into a coherent state due to the photon pressure
when the photon hits the oscillator of rod A. 
Furthermore, in our present analysis, the effective frequency of rod A alters by the second order of optomechanical interaction depending on the photon number; $\omega_{a,0}$ if a single photon hits the rod A, while $\omega_{a,1}$ if not.

As a result, we found two effects that amplify the gravitational signal in the visibility of the quantum optomechanical system. First, we showed that the higher-order contribution of $\theta_a$ makes the visibility further sensitive to the quantum gravity effect.
The gravitational modification in the visibility was given by the first order of the gravitational coupling $g$, while it appears from the second order of $g$ in the previous works~\cite{Balushi2018,Matsumura2020}.
Another way to enhance the quantum gravity signal is to make use of the resonance. Since the setup contains two oscillators A and B, the resonance occurs when the two frequencies of these oscillators are close enough. 
We also found the relational equation between the visibility and negativity. This reveals that the resonance effect occurs both in the visibility and negativity at the same time. 

By combining the two effects found in this work, we expect to improve the quantum gravity signal significantly in an optomechanical experiment, which may lead to the implementation of the quantum Cavendish experiment in the near future. 
However, there are still some difficulties preventing a sufficient profit in our approach. 
In our analysis, the characteristic times of the two enhancements are given by
$t\approx (\omega_{a,1}-\omega_{a,0})^{-1}$ and 
$t\approx (\omega_{b}-\omega_{a,1})^{-1}$, respectively. 
The frequency difference between $\omega_{a,0}$ and $\omega_{a,1}$ is typically tiny, and the large resonance enhancement is realized for the small matching parameter $\epsilon:=1-\omega_{b}/\omega_{a,1}$. 
Hence, for utilizing the two enhancements, we need to coherently sustain our system for a long time, and this may be a challenging issue.
Despite that, our investigation gives remarkable suggestions to enhance the quantum gravity signal in the conventional experimental setup. Particularly, the resonance effect can be very useful not only in our setup but also in many systems containing several oscillators.

\begin{acknowledgements}
We thank Nobuyuki Matsumoto for useful
discussions related to the topic of this paper.
This work is supported by the Nagoya University Interdisciplinary Frontier Fellowship (Y.K.), JSPS KAKENHI grants 20H05854, 23K03424 (T.F.), and 23K13103 (A.M.).
\end{acknowledgements}

\appendix

\section{Gravitational interaction Hamiltonian}
\label{apdx:Hamiltonian}

We will show how to obtain the gravitational interaction Hamiltonian in Eq.~\eqref{eq:total_Hamiltonian}.
We first assume $1\gg h/L\gg \theta_b-\theta_a$. This assumption indicates that the vertical separation of two rods is much smaller than the length of each rod to focus on gravity mediating only between mirrors located near each other. Also, the oscillation of rods is negligible compared to the vertical separation of rods. Considering a quantized form of Newtonian gravity between mirrors of rod A and B with the above assumption,  we get
\begin{align}
    \frac{-2G m M}{\sqrt{h^2+\left(2L\sin[(\hat\theta_b-\hat\theta_a)/2]\right)^2}}
    \approx \frac{GmML^2}{h^3}\left(\hat\theta_a^2+\hat\theta_b^2-2\hat\theta_a\hat\theta_b\right),
    \label{eq:Hg}
\end{align}
where we neglected a constant term.  
The first and the second terms in the last line play a role to shift the original oscillation frequency of each rod. The last term coupling angular positions of two rods induces gravity-induced entanglement between them.
We inserted this expression to the first line of Eq.~\eqref{eq:total_Hamiltonian}.

\section{Time evolved state}
\label{apdx:time_evolution}

Here, we derive a time evolution of the total state given in Eq.~\eqref{eq:time_evolution} and show its explicit form.
Using the Hamiltonian in Eq.~\eqref{eq:total_Hamiltonian} and the initial state in Eq.~\eqref{eq:initial_state}, the time evolved state is given by
\begin{align}
    |\psi(t)\rangle
    =\frac{e^{-i\omega_c t}}{\sqrt{2}}\sum_{n=0,1}
    \ket{n,1-n}_c
    e^{-i\left(\hat H_{a,n}+\hat H_b+\hat H_g\right)t/\hbar}
    \ket{\alpha}_a \ket{\beta}_b
\end{align}
In the following, we focus on the state of rod A and B written as $e^{-i\left(\hat H_{a,n}+\hat H_b+\hat H_g\right)t/\hbar}\ket{\alpha}_a \ket{\beta}_b$.
First, we move on to the interaction picture and consider up to the first order of gravitational coupling constant $g$.
We denote the free evolution Hamiltonian without gravity and gravitational interacting Hamiltonian as follows.
\begin{align}
    \hat H_n^{(0)}=\hat H_{a,n}+\hat H_b,\quad
    \hat H_{g,n}^I(t):=e^{i\hat H_n^{(0)}t/\hbar}\hat H_g e^{-i\hat H_n^{(0)}t/\hbar}
\end{align}
Using these Hamiltonians, the time-evolved state of rods A and B is rewritten as
\begin{align}
    e^{-i\hat H_n^{(0)}t/\hbar}\ket{\alpha}_a \ket{\beta}_b
    &=e^{-i\hat H_n^{(0)}t/\hbar}~
    \mathcal{T}\left[
    \exp\left[-\frac{i}{\hbar}\int_0^t dt' \hat H_{g,n}^I(t')\right]
    \right]\ket{\alpha}_a \ket{\beta}_b\notag\\
    &\approx e^{-i\hat H_n^{(0)}t/\hbar}\left(
    1-\frac{i}{\hbar}\int_0^t dt' \hat H_{g,n}^I(t')
    \right)\ket{\alpha}_a \ket{\beta}_b
    +\mathcal{O}[g^2].
\end{align}
In the second line, we take into account the first order of $g$. 
By using the following relation satisfied for the interaction picture Hamiltonian
\begin{align}
    e^{-i\hat H_n^{(0)}t/\hbar}H_{g,n}^I(t')
    =H_{g,n}^I(t'-t)e^{-i\hat H_n^{(0)}t/\hbar},
\end{align}
we obtain the time-evolved state as
\begin{align}
    e^{-i\hat H_n^{(0)}t/\hbar}\ket{\alpha}_a \ket{\beta}_b
    &\approx \left(
    1-\frac{i}{\hbar}\int_0^t dt' \hat H_{g,n}^I(t'-t)
    \right)
    e^{-i\hat H_{a,n}t/\hbar}\ket{\alpha}_a 
    e^{-i\hat H_bt/\hbar}\ket{\beta}_b.
    \label{eq:stateAB}
\end{align}

Next, we investigate the explicit form of the free evolution state of rod A, $e^{-i\hat H_{a,n}t/\hbar}\ket{\alpha}_a$, contained in Eq.~\eqref{eq:stateAB}. 
Beforehand, we should note that the initial coherent state $|\alpha\rangle_a$ is a coherent eigenstate of $\hat a_0$, but not of $\hat a_1$. As we see in the following, $|\alpha\rangle_a$ is regarded as a squeezed coherent state in terms of $\hat a_1$.
The relationship between $\hat a_0$ and $\hat a_1$ is given by
\begin{gather}
    \hat a_1
    =
    \hat S[\zeta_1] \hat a_0 \hat S^\dagger[\zeta_1]
    =\cosh[\zeta_1]\hat a_0+\sinh[\zeta_1]\hat a_0^\dagger,\\
    \zeta_n:=-\frac{1}{2}\log \left[\frac{\omega_{a,0}}{\omega_{a,n}}\right],\qquad
    \hat S[\xi]:=\exp\left[\frac{1}{2}\left(\xi^*\hat a_0^2-\xi \hat a_0^2\right)\right]
    =\exp\left[\frac{1}{2}\left(\xi^*\hat a_1^2-\xi \hat a_1^2\right)\right].
\end{gather}
Here, $\hat S$ is a squeezing operator and $\zeta_n$ is a squeezing parameter. This leads to another relative equation combining two vacuum states of $\hat a_0$ and $\hat a_1$.
\begin{align}
    |0\rangle_{a,0}=\hat S[-\zeta_1]|0\rangle_{a,1},\qquad
    \text{where}
    \quad
    \hat a_0|0\rangle_{a,0}=\hat a_1|0\rangle_{a,1}=0
\end{align}
Furthermore, the above equation is extended to a relative equation connecting a coherent state of $\hat a_0$ to another state of $\hat a_1$.
\begin{gather}
    |\alpha\rangle_a=|\alpha\rangle_{a,0}
    =\hat D_n \left[\alpha_n\right]\,\hat S\left[-\zeta_n\right]|0\rangle_{a,n}
    =|\alpha,-\zeta_1\rangle_{a,1},
    \label{eq:initial_stateA}\\
    \hat D_n[\eta]:=\exp\left[\eta^*\hat a_n^\dagger-\eta \hat a_n\right],\qquad
    \alpha_n:=\cosh\left[\zeta_n\right]\alpha+\sinh\left[\zeta_n\right]\alpha^*,\qquad
    |\eta,\xi\rangle_{a,n}=\hat D_n \left[\eta\right]\,\hat S\left[\xi\right]|0\rangle_{a,n}
\end{gather}
Here, $\hat D_n$ and $\alpha_n$ are the displacement operator and the coherent parameter defined in terms of $\hat a_n$ respectively. $|\eta,\xi\rangle_{a,n}$ is the squeezed coherent state concerning $\hat a_n$. This relational equation indicates that the initial coherent state of $\hat a_0$ is equivalent to a squeezed coherent state of $\hat a_1$. 
Since the Hamiltonian of rod A contains both $\hat a_0$ and $\hat a_1$, we need to solve time evolution for the squeezed coherent state in general. The squeezing effect in our calculation arises from the fact that we consider the higher-order optomechanical contribution $\mathcal{O}[\theta_a^2]$, two different frequencies $\omega_{a,0},~\omega_{a,1}$ were introduced, and two different annihilation operators $\hat a_0,~\hat a_1$ appears in the Hamiltonian.

The free time evolution operator of rod A is rewritten as follows.
\begin{align}
    e^{-i\hat H_{a,n}t/\hbar}
    =e^{i\phi'_n}\hat D_n\left[n\lambda_n\right]
    \exp\left[-i\omega_{a,n}t\hat a_n^\dagger\hat a_n\right]
    \hat D_n^\dagger\left[n\lambda_n\right],\quad
    \phi'_n:=\omega_{a,n}\left(n\lambda_n^2-\frac{1}{2}\right)t
    \label{eq:time_evolutionA}
\end{align}
This expression clearly shows that the original harmonics oscillator potential $e^{-i\omega_{a,n}t\hat a_n^\dagger\hat a_n}$ is shifted horizontally with the coherent parameter $n \lambda_n$.
Combining Eq.~\eqref{eq:initial_stateA} and \eqref{eq:time_evolutionA}, the free evolution state of rod A is given as
\begin{align}
    e^{-i\hat H_{a,n}t/\hbar}|\alpha\rangle_{a,0}
    &=e^{i\phi'_n}\hat D_n\left[n\lambda_n\right]\exp\left[-i\omega_{a,n}t\hat a_n^\dagger\hat a_n\right]\hat D_n^\dagger\left[n\lambda_n\right]
    \hat D_n \left[\alpha_n\right]\,\hat S\left[-\zeta_n\right]|0\rangle_{a,n}\notag\\
    &=e^{i\phi_n}\left|\Phi_{a,n},e^{-2i\omega_{a,n}t}\zeta_n\right\rangle_{a,n},
    \label{eq:stateA}
\end{align}
where
\begin{align}
    \phi_n:=\phi'_n
    +n\lambda_n\left\{
    \mathrm{Im}\left[\alpha_{a,n}\left(1-e^{-i\omega_{a,n}t}\right)\right]
    -\lambda_n\sin[\omega_{a,n}t]
    \right\},\qquad
    \Phi_{a,n}:=e^{-i\omega_{a,n}t}\alpha_n+n\lambda_n\left(1-e^{-i\omega_{a,n}t}\right).
\end{align}
From the first line to the second line in Eq.~\eqref{eq:stateA}, we make use of the following relation.
\begin{align}
    &e^{-i\omega_{a,n}t\hat a_n^\dagger\hat a_n}\hat D_n\left[\eta\right]
    =\hat D_n\left[e^{-i\omega_{a,n} t}\eta\right]e^{-i\omega_{a,n}t\hat a_n^\dagger\hat a_n},\qquad
    e^{-i\omega_{a,n}t\hat a_n^\dagger\hat a_n}\hat S\left[\xi\right]
    =\hat S\left[e^{-2i\omega_{a,n}t}\xi\right]e^{-i\omega_{a,n}t\hat a_n^\dagger\hat a_n}
\end{align}

With a similar calculation, we obtain the free time evolution of rod B as follows.
\begin{align}
    e^{-i\hat H_bt/\hbar}|\beta\rangle_b=|\Phi_b\rangle,\quad
    \Phi_b:=e^{-i\omega_b t}\beta
    \label{eq:stateB}
\end{align}

Next, we show the explicit form of the gravitational interacting part in the time evolution operator $\frac{i}{\hbar}\int^t_0 dt' \hat H_{g,n}^I(t'-t)$. 
First, we rewrite the gravitational interacting Hamiltonian in the context of $\hat a_n$.
\begin{align}
    \hat H_g:=-g \hbar\omega_{a,0} (\hat a^\dagger_0+\hat a_0)(\hat b^\dagger+\hat b)
    =-g \hbar\sqrt{\omega_{a,n}\omega_{a,0}} (\hat a^\dagger_n+\hat a_n)(\hat b^\dagger+\hat b)
\end{align}
Using the above expression and Eq.~\eqref{eq:time_evolutionA}, we get
\begin{align}
    \hat H_{g,n}^I(t)
    &=-g \hbar\sqrt{\omega_{a,n}\omega_{a,0}}~
    \hat D_n\left[n\lambda_n\right]
    e^{i\omega_{a,n}t\hat a_n^\dagger\hat a_n}
    \hat D_n^\dagger\left[n\lambda_n\right]
    (\hat a^\dagger_n+\hat a_n)
    \hat D_n\left[n\lambda_n\right]
    e^{-i\omega_{a,n}t\hat a_n^\dagger\hat a_n}
    \hat D_n^\dagger\left[n\lambda_n\right]\notag\\
    &\hspace{200pt}\otimes  e^{i\hat H_bt/\hbar}(\hat b^\dagger+\hat b)e^{-i\hat H_bt/\hbar}\notag\\
    &=-g \hbar\sqrt{\omega_{a,n}\omega_{a,0}}~
    \left[
    e^{i\omega_{a,n}(t)}\hat a_{n}^\dagger+e^{-i\omega_{a,n}(t)}\hat a_{n}+2n\lambda_{a,n}(1-\cos(\omega_{a,n}(t)))
    \right]\notag\\
    &\hspace{200pt}\otimes\left[
    e^{i\omega_b(t)}\hat b^\dagger+e^{-i\omega_b(t)}\hat b
    \right]
\end{align}
Finally, by integrating this interaction picture Hamiltonian, we obtain 
\begin{align}
    \frac{i}{\hbar}\int^t_0 dt' \hat H_{g,n}^I(t'-t)
    &=-i g \sqrt{\omega_{a,n}\omega_{a,0}}
    \int^t_0 dt'\left[
    e^{i\omega_{a,n}(t'-t)}\hat a_{n}^\dagger+e^{-i\omega_{a,n}(t'-t)}\hat a_{n}+2n\lambda_{a,n}(1-\cos(\omega_{a,n}(t'-t)))
    \right]\notag\\
    &\hspace{200pt}\otimes\left[
    e^{i\omega_b(t'-t)}\hat b^\dagger+e^{-i\omega_b(t'-t)}\hat b
    \right]\notag\\
    &=-2i g \left(\hat{\mathcal{I}}_n(t)+n \hat{\mathcal{J}}(t)\right).
    \label{eq:time_evolution g}
\end{align}
Here $\hat{\mathcal{I}}_n(t),~\hat{\mathcal{J}}(t)$ is defined in Eq.~\eqref{eq:I}\,\eqref{eq:J}.

Finally, according to Eq.~\eqref{eq:stateA}\,\eqref{eq:stateB} and \eqref{eq:time_evolution g}, the time evolved state is
\begin{align}
    |\psi(t)\rangle
    =\frac{e^{-i\omega_c t}}{\sqrt{2}}\sum_{n=0,1}
    \ket{n,1-n}_c
    \left(
    1+2i g \left(\hat{\mathcal{I}}_n(t)+n \hat{\mathcal{J}}(t)\right)
    \right)
    e^{-i\hat H_{a,n}t/\hbar}\ket{\alpha}_a 
    e^{-i\hat H_bt/\hbar}\ket{\beta}_b
    \label{eq:total_state}
\end{align}
where free evolution states of rod A and B are given by
\begin{align}
    e^{-i\hat H_{a,n}t/\hbar}\ket{\alpha}_a
    =e^{i\phi_n}\left|\Phi_{a,n},e^{-2i\omega_{a,n}t}\zeta_n\right\rangle_{a,n},\qquad
    e^{-i\hat H_bt/\hbar}|\beta\rangle_b=|\Phi_b\rangle.
\end{align}

\section{The explicit form of the visibility}
\label{apdx:visibility}

In this section, we demonstrate an explicit form of visibility in Eq.~\eqref{eq:visibility}.
First, we rewrite the 0th order visibility $\mathcal{V}_c^{(0)}(t)$ defined in Eq.~\eqref{eq:visibility result} using Eq.~\eqref{eq:stateA}.
\begin{align}
    \mathcal{V}_c^{(0)}(t)
    &:=\left|_a\langle\alpha|e^{i\hat H_{a,0}t/\hbar}\,e^{-i\hat H_{a,1}t/\hbar}|\alpha\rangle_a\right|\\
    &={}_{a,0}\langle\Phi_{a,0}|
    \Phi_{a,1},e^{-2i\omega_{a,1}t}\zeta_1\rangle_1
    ={}_{a,1}\langle\tilde{\Phi}_{a,0},-\zeta_1|
    \Phi_{a,1},e^{-2i\omega_{a,1}t}\zeta_1\rangle_1\notag
\end{align}
In the third equality, we rewrite the bra state in terms of $\hat a_1$ in a similar way as Eq.~\eqref{eq:initial_stateA}, where its coherent parameter is calculated as
\begin{align}
    \tilde{\Phi}_{a,0}=\cosh[\zeta_1]\Phi_{a,0}+\sinh[\zeta_1]\Phi_{a,0}^*.
\end{align}
We can see that $\mathcal{V}_c^{(0)}(t)$ is an inner product of two squeezed coherent states, and its explicit form is given by
\begin{align}
    \mathcal{V}_c^{(0)}(t)
    &=\left(\frac{\mathrm{Re}[A[e^{-2i\omega_{a,1}t}\zeta_1]]\mathrm{Re}[A^*[-\zeta_1]]}{\pi^2}\right)^{1/4}
    \sqrt{\frac{2\pi}{A[e^{-2i\omega_{a,1}t}\zeta_1]+A^*[-\zeta_1]}}\notag\\
    &\hspace{20pt}\times
    \exp\left[-\mathrm{Re}[A[e^{-2i\omega_{a,1}t}\zeta_1]]\mathrm{Re}[\Phi_{a,1}]^2-\mathrm{Re}[A^*[-\zeta_1]]\mathrm{Re}[\tilde{\Phi}_{a,0}]^2\right]\notag\\
    &\hspace{20pt}\times
    \exp\left[\frac{\left(A[e^{-2i\omega_{a,1}t}\zeta_1]\mathrm{Re}[\Phi_{a,1}]-A^*[-\zeta_1]\mathrm{Re}[\tilde{\Phi}_{a,0}]+i\left(\mathrm{Im}[\Phi_{a,1}]-\mathrm{Im}[\tilde{\Phi}_{a,0}]\right)\right)^2}{A[e^{-2i\omega_{a,1}t}\zeta_1]+A^*[-\zeta_1]}\right]
    \label{eq:visibility0}
\end{align}
where
\begin{align}
    A[\xi]:=\frac{1+(\xi/|\xi|)\tanh[|\xi|]}{1-(\xi/|\xi|)\tanh[|\xi|]}.
\end{align}

Next, we focus on the gravitational contribution to visibility. In advance, the inner product of $\hat a$ using a general coherent squeezed state is given as follows.
\begin{align}
    &\langle \eta',\xi'|\hat a|\eta,\xi\rangle
    =\mathcal{E}\left[\eta',\xi'\right|\left.\eta,\xi\right]\langle \eta',\xi'|\eta,\xi\rangle,\\
    &\mathcal{E}\left[\eta',\xi'\right|\left.\eta,\xi\right]
    :=\frac{(1+A^*[\xi'])(A[\xi] \mathrm{Re}[\eta]+i\,\mathrm{Im}[\eta])+(1-A[\xi])(A^*[\xi']\mathrm{Re[\eta']-i\,\mathrm{Im}[\eta']})}{A[\xi]+A^*[\xi']}
\end{align}
Also, the inner product of $\hat a^\dagger$ is given by
\begin{align}
    &\langle \beta',\zeta'|\hat a^\dagger|\beta,\zeta\rangle
    =\mathcal{E}^*[\beta,\zeta,\beta',\zeta']\langle \beta',\zeta'|\beta,\zeta\rangle.
\end{align}
Then, the inner products of the annihilation and creation operators of rod A appearing in the visibility are given by
\begin{align}
    {}_0\langle\hat a_1\rangle_1
    &:=\frac{_a\langle\alpha|_b\langle\beta|e^{i\left(\hat H_{a,0}+\hat H_b\right)t/\hbar}
    \,\hat a_1\,
    e^{-i\left(\hat H_{a,1}+\hat H_b\right)t/\hbar}|\alpha\rangle_a|\beta\rangle_b}
    {_a\langle\alpha|e^{i\hat H_{a,0}t/\hbar}\,e^{-i\hat H_{a,1}t/\hbar}|\alpha\rangle_a}\notag\\
    &=\frac{{}_{a,1}\langle\tilde{\Phi}_{a,0},-\zeta_1|
    \,\hat a_1\,
    |\Phi_{a,1},e^{-2i\omega_{a,1}t}\zeta_1\rangle_1}
    {{}_{a,1}\langle\tilde{\Phi}_{a,0},-\zeta_1|
    \Phi_{a,1},e^{-2i\omega_{a,1}t}\zeta_1\rangle_1}
    =\mathcal{E}\left[\tilde{\Phi}_{a,0},-\zeta_1\left|\right.
    \Phi_{a,1},e^{-2i\omega_{a,1}t}\zeta_1\right],
    \label{eq:a1}\\
    {}_0\langle\hat a_1^\dagger\rangle_1
    &:=\frac{_a\langle\alpha|_b\langle\beta|e^{i\left(\hat H_{a,0}+\hat H_b\right)t/\hbar}
    \,\hat a_1^\dagger\,
    e^{-i\left(\hat H_{a,1}+\hat H_b\right)t/\hbar}|\alpha\rangle_a|\beta\rangle_b}
    {_a\langle\alpha|e^{i\hat H_{a,0}t/\hbar}\,e^{-i\hat H_{a,1}t/\hbar}|\alpha\rangle_a}\notag\\
    &=\frac{{}_{a,1}\langle\tilde{\Phi}_{a,0},-\zeta_1|
    \,\hat a_1^\dagger\,
    |\Phi_{a,1},e^{-2i\omega_{a,1}t}\zeta_1\rangle_1}
    {{}_{a,1}\langle\tilde{\Phi}_{a,0},-\zeta_1|
    \Phi_{a,1},e^{-2i\omega_{a,1}t}\zeta_1\rangle_1}
    =\mathcal{E}^*\left[\Phi_{a,1},e^{-2i\omega_{a,1}t}\zeta_1\right|\left.
    \tilde{\Phi}_{a,0},-\zeta_1\right],
    \label{eq:a1dagger}\\
    {}_0\langle\hat a_0\rangle_1
    &:={}_0\langle 
    \cosh[\zeta_1]\,\hat a_1
    -\sinh[\zeta_1]\,\hat a_1^\dagger\rangle_1
    =\cosh[\zeta_1]\,{}_0\langle\hat a_1\rangle_1
    -\sinh[\zeta_1]\,{}_0\langle\hat a_1^\dagger\rangle_1
    \label{eq:a0}\\
    {}_0\langle\hat a_0^\dagger\rangle_1
    &:={}_0\langle
    -\sinh[\zeta_1]\,\hat a_1
    +\cosh[\zeta_1]\,\hat a_1^\dagger\rangle_1
    =-\sinh[\zeta_1]\,{}_0\langle\hat a_1\rangle_1
    +\cosh[\zeta_1]\,{}_0\langle\hat a_1^\dagger\rangle_1
    \label{eq:a0dagger}
\end{align}
Similarly, the inner products of rod B operators are given as follows.
\begin{align}
    {}_0\langle\hat b\rangle_1
    =\Phi_b,\qquad
    {}_0\langle\hat b^\dagger\rangle_1
    =\Phi_b^*
\end{align}
Based on these equations, we obtain the inner product of $\hat{\mathcal{I}}_n$ as
\begin{align}
    _0\langle\hat{\mathcal{I}}_n(t)\rangle_1
    &=\sqrt{\frac{\omega_{a,0}^3}{\omega_{a,n}}}
    \left\{\frac{\sin[\omega_{n,+}t/2]}{\omega_{n,+}}\left(
    e^{-i\omega_{n,+}t/2}
    \Phi_b^* ~ {}_0\langle\hat a_{n}^\dagger\rangle_1
    +e^{i\omega_{n,+}t/2}
    \Phi_b ~ {}_0\langle\hat a_{n}\rangle_1
    \right)\right.\notag\\
    &\hspace{60pt}\left.+\frac{\sin[\omega_{n,-}t/2]}{\omega_{n,-}}\left(
    e^{-i\omega_{n,-}t/2}
    \Phi_b ~ {}_0\langle\hat a_{n}^\dagger\rangle_1
    +e^{i\omega_{n,-}t/2}
    \Phi_b^* ~ {}_0\langle\hat a_{n}\rangle_1
    \right)\right\}
    \label{eq:innerproduct_I}
\end{align}
where the expressions of ${}_0\langle\hat a_n\rangle_1,~{}_0\langle\hat a_n^\dagger\rangle_1$ are shown in Eq.~\eqref{eq:a1}-\eqref{eq:a0dagger}.

At last, by substituting Eq.~\eqref{eq:innerproduct_I} into Eq.~\eqref{eq:visibility}, we obtain the final expression for the visibility.
\begin{align}
    \mathcal{V}_c(t)
    &= \mathcal{V}_c^{(0)}(t)
    \left(1
    +2g ~\mathrm{Im}\left[
    {}_0\langle\hat{\mathcal{I}}_0^\dagger(t)\rangle_1-{}_0\langle\hat{\mathcal{I}}_1(t)\rangle_1\right]
    \right)
    +\mathcal{O}[g^2]\\
    &\approx\mathcal{V}_c^{(0)}(t)\left\{1+
    2g\,\omega_{a,0}\left(
    \frac{\sin[\omega_{0,+}t/2]}{\omega_{0,+}}D_{0,+}
    +\frac{\sin[\omega_{1,+}t/2]}{\omega_{1,+}}D_{1,+}\right.\right.
    \notag\\
    &\hspace{160pt}\left.\left.+\frac{\sin[\omega_{0,-}t/2]}{\omega_{0,-}}D_{0,-}
    +\frac{\sin[\omega_{1,-}t/2]}{\omega_{1,-}}D_{1,-}
    \right)\right\}
\end{align}
Here, the coefficient of each term is given by
\begin{align}
    &D_{0,\pm}
    =\sqrt{\frac{\omega_{a,0}}{\omega_{a,1}}}
    \mathrm{Re}\left[e^{\mp i\omega_{a,0}t/2}\beta\right]
    \mathrm{Im}\left[{}_0\langle \hat{a_1}\rangle_1+{}_0\langle \hat{a_1}^\dagger\rangle_1\right]
    \mp \sqrt{\frac{\omega_{a,1}}{\omega_{a,0}}}
    \mathrm{Im}\left[e^{\mp i\omega_{a,0}t/2}\beta\right]
    \mathrm{Re}\left[{}_0\langle \hat{a_1}\rangle_1-{}_0\langle \hat{a_1}^\dagger\rangle_1\right]\\
    &D_{1,\pm}
    =-\sqrt{\frac{\omega_{a,1}}{\omega_{a,0}}}\left(
    \mathrm{Re}\left[e^{\pm i\omega_{a,1}t/2}\beta\right]
    \mathrm{Im}\left[{}_0\langle \hat{a_1}\rangle_1+{}_0\langle \hat{a_1}^\dagger\rangle_1\right]
    \pm
    \mathrm{Im}\left[e^{\pm i\omega_{a,1}t/2}\beta\right]
    \mathrm{Re}\left[{}_0\langle \hat{a_1}\rangle_1-{}_0\langle \hat{a_1}^\dagger\rangle_1\right]
    \right),
\end{align}
and $\mathcal{V}_c^{(0)}(t)$ is given in Eq.~\eqref{eq:visibility0}.
We see that this explicit form reduces to Eq.~\eqref{eq:visibility2} when $\beta$ is a real number.

\section{Negativity between the rod B and other systems}
\label{apdx:entanglement}

Here, we show the derivation of the negativity between rod B and other systems in section\,\ref{sec:entanglement}, and display its explicit form.

We need to get a density matrix of the state. To do so, we define the unit orthogonal bases of each state to construct the matrix. 
Since there are only two kinds of state $|\Phi_b\rangle$ and $\hat b^\dagger |\Phi_b\rangle$ for rod B state in Eq.~\eqref{eq:total_state}, the bases for rod B system is given by two orthogonal states.
\begin{align}
    |b_0\rangle:=|\Phi_b\rangle,\quad
    |b_1\rangle:=\hat b^\dagger |\Phi_b\rangle-\Phi_b^*|\Phi_b\rangle
\end{align}
Then, the time-evolved state is rewritten as
\begin{align}
    |\psi(t)\rangle=\frac{1}{\sqrt{2}}e^{-i\omega_c t}\left(|\psi_0\rangle|b_0\rangle+|\psi_1\rangle|b_1\rangle\right),
\end{align}
where $|\psi_j\rangle$ is the state of rod A and the photon systems
\begin{align}
    &|\psi_0\rangle=|0\rangle\left\{1+2ig\left(\Phi_b\hat{\mathcal{K}}_0
    +\Phi_b^*\hat{\mathcal{K}}_0^\dagger\right)\right\}|\Phi_{a,0}\rangle_{a,0}
    +|1\rangle\left\{1+2i\gamma\left(\Phi_b \hat{\mathcal{K}}_1+\Phi_b^* \hat{\mathcal{K}}_1^\dagger\right)\right\}|\Phi_{a,1},e^{-2i\omega_{a,1}t}\zeta_1\rangle_{a,1},\\
    &|\psi_1\rangle=2i g \left(|0\rangle\,\hat{\mathcal{K}}_0^\dagger|\Phi_{a,0}\rangle_{a,0}+|1\rangle\,\hat{\mathcal{K}}_1^\dagger|\Phi_{a,1},e^{-2i\omega_{a,1}t}\zeta_1\rangle_{a,1}\right).
\end{align}
$\hat{\mathcal{K}}_n$ is defined in Eq.~\eqref{eq:K}.
We also introduce unit orthogonal bases for the complement system of rod B $|\psi_0\rangle,~|\psi_1\rangle$.
\begin{align}
    |\bar{b}_0\rangle:=|\psi_0\rangle,\quad
    |\bar{b}_1\rangle:=\frac{1}{\sqrt{N_{\bar{b}}}}\left(|\psi_1\rangle-\langle\psi_0|\psi_1\rangle|\psi_0\rangle\right)
\end{align}
$\sqrt{N_{\bar{b}}}$ is a normalization factor given by
\begin{align}
    &\sqrt{N_{\bar{b}}}
    =2g\sqrt{\sum_{n=0,1} {}_a\langle\alpha|
    e^{i\hat H_{a,n}t/\hbar}\hat{\mathcal{K}}^{\dagger}_n(t)
    \hat{\mathcal{K}}_n(t)e^{-i\hat H_{a,n}t/\hbar}|\alpha\rangle_a}\,.
    \label{eq:Nb}
\end{align}

Using the bases introduced above, we construct the density matrix.
\begin{align}
    \rho(t)=|\psi(t)\rangle\langle\psi(t)|
    =\sum_{I,J=0}^3 \left(\rho^{(0)}_{IJ}+ \rho^{(1)}_{IJ}\right)|e_I\rangle\langle e_J|
    \label{eq:densitymatrix_total}
\end{align}
$\rho^{(0)}$ and $\rho^{(1)}$ are the density matrix of the 0th order and the first order of $g$.
$|e_J\rangle$ is the composite bases of the total system
\begin{align}
    |e_0\rangle=|b_0\rangle|\bar{b}_0\rangle,\quad
    |e_1\rangle=|b_1\rangle|\bar{b}_0\rangle,\quad
    |e_2\rangle=|b_0\rangle|\bar{b}_1\rangle,\quad
    |e_3\rangle=|b_1\rangle|\bar{b}_1\rangle,\quad
\end{align}
and the matrix components are given by
\begin{align}
    \rho^{(0)}=
    \begin{pmatrix}
        1 &0&0&0\\
        0&0&0&0\\
        0&0&0&0\\
        0&0&0&0
    \end{pmatrix},
    \quad
    \rho^{(1)}=
    \begin{pmatrix}
        0&0&\langle\psi_1|\psi_0\rangle&\sqrt{N_{\bar{b}}}\\
        0&0&0&0\\
        \langle\psi_0|\psi_1\rangle&0&0&0\\
        \sqrt{N_{\bar{b}}}&0&0&0
    \end{pmatrix}
    \,.
\end{align}
Then, we perform a partial transpose to the density matrix, solve its eigenvalues up to the first order of $g$. Note that $\rho^{(0)}$ is triple-degenerated, so we need to solve a degenarated eigensystem. 
Finally, by estimating a total sum of the negative eigenvalues, we find that the negativity is given by the normalization factor $\sqrt{N_{\bar{b}}}$.
\begin{align}
    &\mathcal{N}_{\text{B:others}}=\sqrt{N_{\bar{b}}}\\
    &=2 g \omega_{a,0} \left[
    \left(\frac{\sin[\omega_{0,+}t/2]}{\omega_{0,+}}\right)^2(|\Phi_{a,0}|^2+1)
    +\left(\frac{\sin[\omega_{0,-}t/2]}{\omega_{0,-}}\right)^2|\Phi_{a,0}|^2
    \notag\right.\\
    &\hspace{20pt}\left.+4\frac{\sin[\omega_{0,+}t/2]}{\omega_{0,+}}\frac{\sin[\omega_{0,-}t/2]}{\omega_{0,-}}
    \cos\left[\frac{\omega_b t}{2}\right]
    \,\mathrm{Re}\left[e^{i\omega_a t}\Phi_{a,0}^2\right]
    \notag\right.\\
    &\hspace{20pt}\left.
    +k^2\left\{
    \left(\frac{\sin[\omega_{1,+}t/2]}{\omega_{1,+}}\right)^2(|\Phi_{a,1}|^2+\cosh^2|\zeta_1|)
    +\left(\frac{\sin[\omega_{1,-}t/2]}{\omega_{1,-}}\right)^2(|\Phi_{a,1}|^2+\sinh^2|\zeta_1|)
    \notag\right.\right.\\
    &\hspace{20pt}\left.\left.
    +4\frac{\sin[\omega_{1,+}t/2]}{\omega_{1,+}}\frac{\sin[\omega_{1,-}t/2]}{\omega_{1,-}}
    \cos\left[\frac{\omega_b t}{2}\right]
    \,\mathrm{Re}\left[e^{i\omega_{a,1} t}(\Phi_{a,1}^2-e^{-2i\omega_{a,1}t}\sinh |2\zeta_1|)\right]
    \notag\right.\right.\\
    &\hspace{20pt}\left.\left.
    +2\lambda_1\mathrm{Re}\left[\frac{F(t)}{\omega_b}\left(
    \frac{\sin[\omega_{1,+}t/2]}{\omega_{1,+}}e^{-i\omega_{1,+}t/2}\Phi_{a,1}^*
    +\frac{\sin[\omega_{1,-}t/2]}{\omega_{1,-}}e^{i\omega_{1,-}t/2}\Phi_{a,1}
    \right)\right]
    +\lambda_1^2\left|\frac{F(t)}{\omega_b}\right|^2\right\}
    \right]^{1/2}.
\end{align}
From the first line to the second line, we calculate the inner product in Eq.~\eqref{eq:Nb} using the following formulas
\begin{align}
    &\langle \eta,\xi|\hat a^2|\eta,\xi\rangle
    =\eta^2-e^{i\theta}\sinh 2r,\quad
    \langle \eta,\xi|\hat a^{\dagger2}|\eta,\xi\rangle
    =\eta^{*2}-e^{-i\theta}\sinh 2r,\\
    &\langle \eta,\xi|\hat a \hat a^\dagger|\eta,\xi\rangle
    =|\eta|^2+\cosh^2 r,\quad
    \langle \eta,\xi|\hat a^\dagger \hat a|\eta,\xi\rangle
    =|\eta|^2+\sinh^2 r,
\end{align}
where $\xi=r e^{i\theta}$.

\end{document}